\documentclass[graybox]{svmult}

\usepackage{cite}       %%% I have added this myself for bibteX
\usepackage{mathptmx}       % selects Times Roman as basic font
\usepackage{helvet}         % selects Helvetica as sans-serif font
\usepackage{courier}        % selects Courier as typewriter font
\usepackage{type1cm}        % activate if the above 3 fonts are
\usepackage{makeidx}         % allows index generation
\usepackage{graphicx}        % standard LaTeX graphics tool
\usepackage{multicol}        % used for the two-column index
\usepackage[bottom]{footmisc}% places footnotes at page bottom

\makeindex             % used for the subject index
\begin{document}

\title*{Magnetic Structure of Actinide Metals}

\author{G. van der Laan and K. T. Moore}
\institute{Gerrit van der Laan \at Diamond Light Source, Chilton, Didcot, Oxfordshire OX11 0DE,
United Kingdom, \\ \email{gerrit.vanderlaan@diamond.ac.uk}
\and Kevin T. Moore \at Lawrence Livermore National Laboratory, Livermore, California 94550, USA}

\maketitle

\abstract*{
In comparison to $3d$ or $4f$ metals, magnetism in actinides remains poorly understood due to experimental complications and the exotic behavior of the $5f$ states. In particular, plutonium metal is most especially vexing. Over the last five decades theories proposed the presence of either ordered or disordered local moments at low temperatures. However, experiments such as magnetic susceptibility, electrical resistivity, nuclear magnetic resonance, specific heat, and elastic and inelastic neutron scattering show no evidence for ordered or disordered magnetic moments in any of the six phases of plutonium. Beyond plutonium, the magnetic structure of other actinides is an active area of research given that temperature, pressure, and chemistry can quickly alter the magnetic structure of the $5f$ states. For instance, curium metal has an exceedingly large spin polarization that results in a total moment of $\sim$8 $\mu_{\mathrm{B}}$/atom, which influences the phase stability of the metal.
Insight in the actinide ground state can be obtained from core-level x-ray absorption spectroscopy (XAS) and electron energy-loss spectroscopy (EELS). A sum rule relates the branching ratio of the core-level spectra measured by XAS or EELS to the expectation value of the angular part of the spin-orbit interaction.
}

\abstract{
In comparison to $3d$ or $4f$ metals, magnetism in actinides remains poorly understood due to experimental complications and the exotic behavior of the $5f$ states. In particular, plutonium metal is most especially vexing. Over the last five decades theories proposed the presence of either ordered or disordered local moments at low temperatures. However, experiments such as magnetic susceptibility, electrical resistivity, nuclear magnetic resonance, specific heat, and elastic and inelastic neutron scattering show no evidence for ordered or disordered magnetic moments in any of the six phases of plutonium. Beyond plutonium, the magnetic structure of other actinides is an active area of research given that temperature, pressure, and chemistry can quickly alter the magnetic structure of the $5f$ states. For instance, curium metal has an exceedingly large spin polarization that results in a total moment of $\sim$8 $\mu_{\mathrm{B}}$/atom, which influences the phase stability of the metal.
Insight in the actinide ground state can be obtained from core-level x-ray absorption spectroscopy (XAS) and electron energy-loss spectroscopy (EELS). A sum rule relates the branching ratio of the core-level spectra measured by XAS or EELS to the expectation value of the angular part of the spin-orbit interaction.
}

%\tableofcontents              %  table of contents 
 
\section{Introduction}
\label{sec:introduction}

Actinide elements such as uranium and plutonium often conjure visions of radiation-spawned monsters who wreak havoc on unsuspecting bystanders. While this makes for spectacular pulp fiction, it does not relay the fact that actinide materials are the cornerstone of nuclear energy and are scientifically fascinating. At present, nuclear power offers the only viable carbon-free energy production that can meet the growing demands of the world that will stress the power grid over the next several decades.  Scientifically, the $5f$ electron states of actinide materials create a myriad of exciting physical behaviors, such as superconductivity, itinerant magnetism, electron correlations, and heavy fermions \cite{moore21}. Therefore, these heavyweights of the Periodic Table are relevant both technologically and scientifically, posed to play a key role in the resurgence of nuclear power for clean and robust energy and expand our understanding of condensed-matter physics.
	
The actinide series consists of fourteen elements and, accordingly, the $5f$ states can contain up to the same number of electrons. Adhering to Hund's rule of maximizing total spin, this means the $f$-shell accepts seven electrons with the same spin direction before taking electrons with opposite spin. Therefore, magnetism has the potential to be strong near the middle of the series. Indeed, this is observed in the rare-earth series where the middle elements such as Eu and Gd exhibit very strong magnetic moments due to spin polarization of the $4f$ states. The strong magnetic moments observed in the rare-earth series are a direct consequence of the $4f$ states being localized, i.e., reduced wavefunction overlap with non- or weakly-bonding $4f$ electrons.  The opposite is true for the light actinides, such as Th, Pa, U, Np, and $\alpha$-Pu, because the $5f$ states are delocalized in the metallic form.  This changes in the latter part of the actinide series where the $5f$ states become localized. The actinide series, therefore, does not simply adhere to Hund's rule. Rather, there is a competition between the exchange interaction that drives maximization of the total spin and the spin-orbit interaction of the $j$ = 5/2 and 7/2 angular momentum levels of the $5f$ states \cite{vanderlaan04,moore06a,moore07a,moore07b}. This competition produces interesting magnetic behavior in the actinides when the states localize in heavier elements and when the light actinide elements bond with other elements.

In this Chapter, we discuss in general terms the magnetism of the actinide elements, in particular Pu. A comparison will be draw with the $3d$, $4d$, and $5d$ metal series as well as the $4f$ metals. Because electron localization is a requirement to obtain local magnetic moments, we will pay specific attention to this in both metals and some compounds. For instance, the transitions from $\alpha$- to $\delta$-Pu and from $\alpha$- to $\gamma$-Ce both exhibit a large change in volume, indicative for a transition from itinerant to localized behavior. However, unlike Ce metal that shows a magnetic moment in the large-volume $\gamma$-Ce phase with localized $4f$ states, both $\alpha$- to $\delta$-Pu show no evidence of a magnetic moment. Surprises like this, litter the actinide series, often creating controversies that are fodder for fruitful and entertaining arguments. We will specifically discuss subjects such as

\begin{itemize}
\item{Localization of the $5f$ states across series, }
\item{Crystal phase transition and comparison to Cerium, }
\item{Magnetic properties, }
\item{Consequences of angular-momentum coupling of the $5f$ states: $LS$, $jj$, and intermediate coupling, }
\item{Role of spin-orbit and electrostatic interactions, }
\item{Electron energy-loss spectroscopy (EELS) in a transmission electron microscope (TEM), }
\item{$N_{4,5}$ branching ratio, }
\item{Spin-orbit sum rule and its validity for the 5$f$ states of actinides, }
\item{Spin-orbit interaction per hole for Th, U, Np, Pu, Am, and Cm, }
\item{Electron population of the $f_{5/2}$ and $f_{7/2}$ levels, }
\item{Ramifications of these results for the magnetic structure of actinides. }
\end{itemize}

We point the reader to other review articles, such as {\it{Challenges in plutonium science}}, Vol.~I and II in the {\it{Los Alamos Science Series}} \cite{Pu}, particularly the article entitled {\emph{Plutonium Condensed-Matter Physics. A Survey of Theory and Experiment}} by Boring and Smith \cite{boring00}, {\emph{Absence of Magnetic Moments in Plutonium}} by Lashley {\emph{et al.}} \cite{lashley05}, {\it{Handbook on the Physics and Chemistry of the Actinides}} by Freeman and Lander \cite{freeman84}, and {\emph{Nature of the 5f States in Actinide Metals}} by Moore and van der Laan \cite{moore21}. These reviews cover the electronic, magnetic, and crystal structure of the actinide materials.  Here, we will focus on the magnetic structure of actinide metals and materials, incorporating the most current experimental and theoretical results.

\section{Volume Change Across the Actinide Series}
\label{sec:volume}

\begin{figure}[t]        %FIGURE Radius            FIG. 1
\sidecaption[t]
% For example, with the graphicx style use
\includegraphics[scale=0.21]{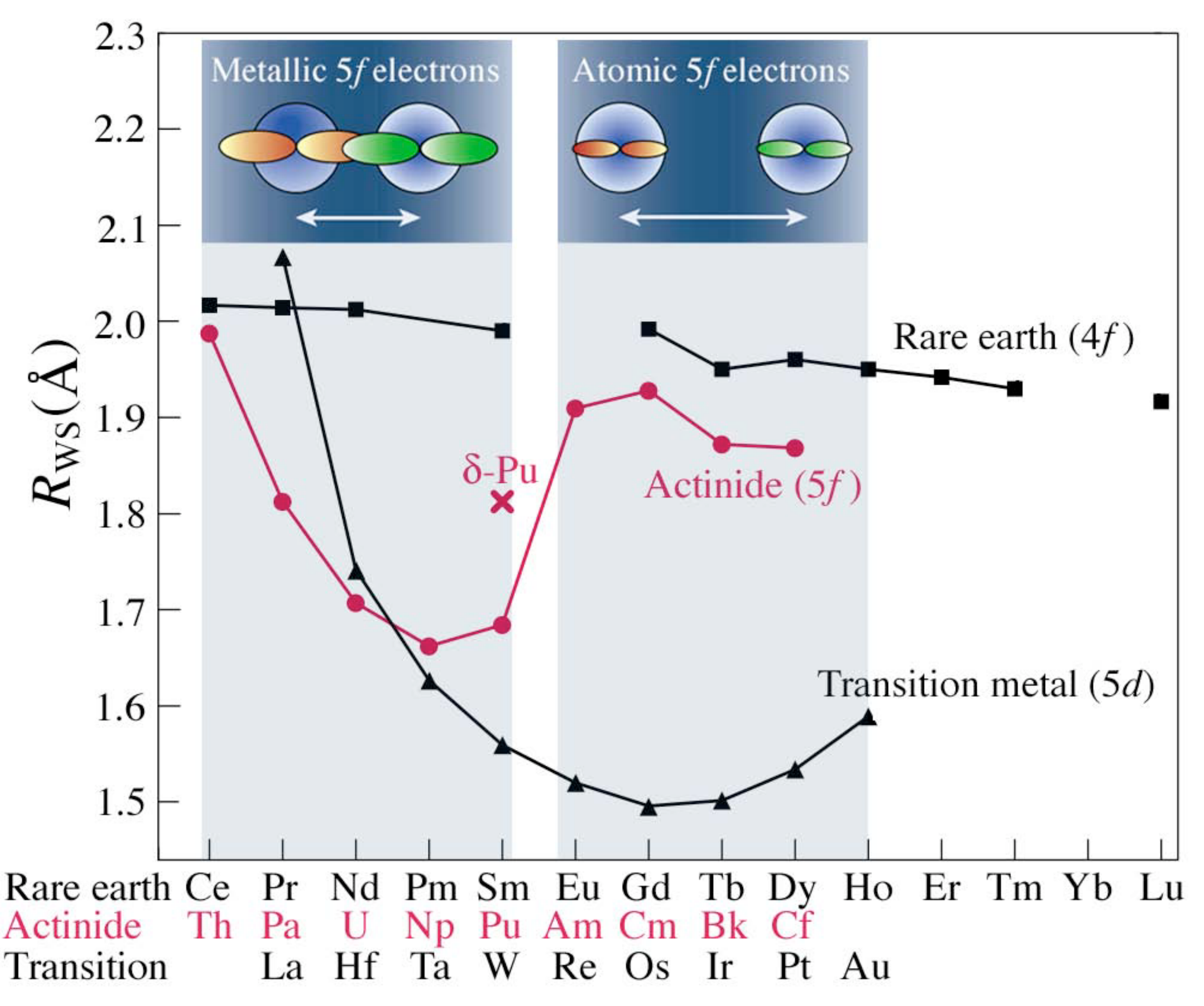}
% If no graphics program available, insert a blank space i.e. use
%\picplace{5cm}{2cm} % Give the correct figure height and width in cm
%
\caption{ Wigner-Seitz radius of each metal as a function of atomic number $Z$ for the $5d$, $4f$, and $5f$ metal series (from Ref.~\cite{boring00}). The upper-left insets schematically illustrate the localized and delocalized $5f$ states between adjacent actinide atoms (from Ref.~\cite{albers01}). Note the parabolic shape of the $3d$ series, the almost constant values of the trivalent rare earths (for clarity the divalent Eu and Yb are omitted), and the unusual behavior of the $5f$ elements, with minimum volume near Pu, and strong increase between Pu and Am. }
\label{Radius}
\end{figure}  %End Figure

Figure \ref{Radius} shows a plot of the metallic radii, which are directly proportional to volume of the actinide $5f$ metals, in comparison to those of $5d$ and $4f$ metals. The curve for the $5d$ transition-metal series is parabolic because in the first half of the series each additional electron contributes to the cohesive energy of the solid, resulting in a {\emph{decrease}} of volume until the shell is approximately half full. In the second half of the $d$ series there is an {\emph{increase}} of volume due to filling of the antibonding states. This behavior is indicative of states that are delocalized and actively bonding. The early actinides (Th to Np) exhibit a {\emph{contraction}} with increasing electron count similar to the $5d$ series, and this in conjunction with the absence of magnetic order indicates that the $5f$ electrons of the early actinide elements are {\emph{itinerant}} (delocalized). 

In the vicinity of Pu there is an abrupt jump in the volume and the elements Am to Cf exhibit no change in volume with increased $5f$ electron count. This behavior can be explained by examining the $4f$ rare-earth series in Fig.~\ref{Radius}. Here, there is no change in metallic radii across the series because the $4f$ states are localized and non-bonding, leaving cohesion to the $(spd)^3$ electrons that do not change count across the series. Comparison with the $4f$ elements, together with the presence of ordered magnetism in Cm and elements beyond, indicate {\it{localized}} behavior of the $5f$ electrons. 

Our discussion of the two types of bonding behaviors in the actinide series emphases the unique position of plutonium. The volume change between $\alpha$-Pu and Am is $\sim$50\%, a staggering change in volume between two neighboring elements in the Periodic Table considering that the only apparent change is to add one electron in the $5f$ shell. Indeed, unlike the lanthanide elements Eu and Yb, which are both divalent in the normally trivalent lanthanide series, there is no indication of a straightforward valence change between Pu and Am. This means that the simple addition of another $5f$ electron is not the pressing issue, but rather the change from itinerant to localized behavior. What we will find is that the bulk of the transition from itinerant to localized $5f$ states occurs within Pu itself! Compared to the ground state $\alpha$-phase, the high-temperature $\delta$-Pu phase is $\sim$25\% larger in volume. Thus, plutonium is the nexus between itinerant- and localized-electron behavior.

\subsection{Photoemission spectroscopy's two cents}

Clear support for increased localization across actinide series is provided by $4f$ core and $5f$ valence-band photoemission of actinide metals. When examining the photoemission spectra for the elements traversing the series, the $4f$ photoemission of $\alpha$-Th  \cite{moser84}, $\alpha$-U \cite{moser84}, and $\alpha$-Np  \cite{naegele87} is dominated by well-screened peaks, characteristic for itinerant behavior. Between $\alpha$-Pu and $\delta$-Pu poorly-screened peaks that are due to more localized $5f$ states increase in intensity compared to the well-screened peak  \cite{arko00}.  $\alpha$-Am shows almost uniquely unscreened peaks, which are characteristic for localized behavior  \cite{naegele84}. In a similar fashion, valence-band photoemission of actinide metals shows an increase in structure from Np to Am, where growing structure indicates localization of the $5f$ states \cite{baer80,naegele84,naegele87,gouder01}. The increase of structure in $4f$ core and $5f$ valence-band photoemission from Np to Am shows that the entire transition from itinerant to localized $5f$ states occurs over several elments, however, the $\alpha$-Pu and $\delta$-Pu spectra clearly reveal that the brunt of the change occurs within the metal's phases.
	
The most important insight from the previous discussion is that the $5f$ states in Th to Pu are slightly overlapping between neighboring actinide atoms. They therefore occupy a very narrow energy band with very high density of states near the Fermi energy. As the number of $5f$ electrons populating the $5f$ band increases, the specific properties of the band begin to dominate the bonding properties of the metal. At plutonium, the $5f$ states rapidly change towards localization and this abrupt change produces a metal with more solid allotropic phases than any other element in the Periodic Table.

\section{The Six Crystal Allotropes  of Pu Metal}
\label{sec:six}

\begin{figure}[t]           %FIGURE Pu-phase           FIG. 2
\sidecaption[t]
\includegraphics[scale=.30]{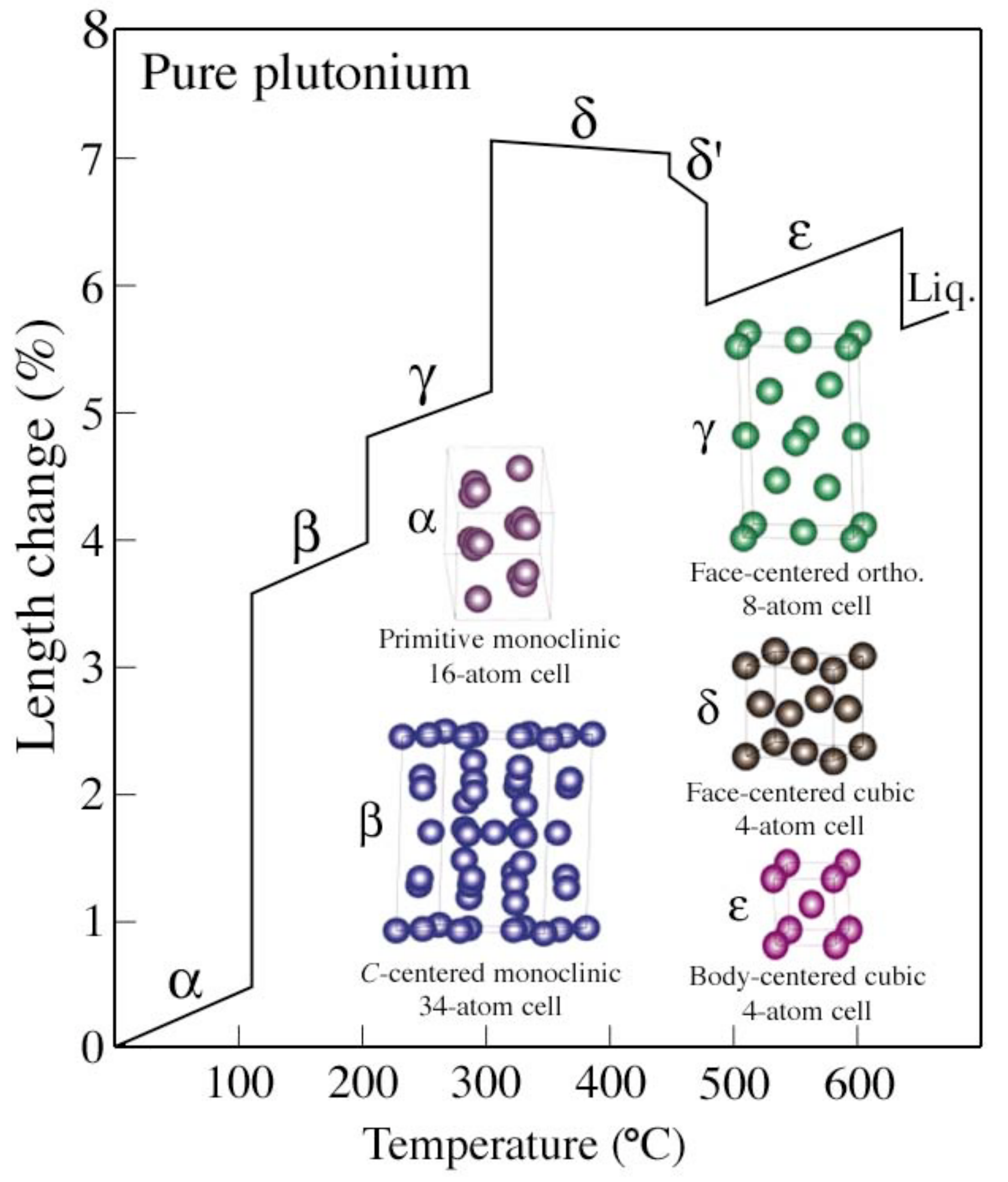}
\caption{ 
Length change of Pu as a function of temperature, including the liquid phase. The crystal structure of all six solid allotropic phases of the metal is given in the lower-right side of the figure (after  Ref.~\cite{hecker00}). The structure changes from low-symmetry monoclinic ($\alpha$-phase) to high-symmetry fcc  ($\delta$-phase), which occurs with exceeding large volume changes over a short temperature span.   Note also the $\delta$-phase contracts as it is heated, and plutonium contracts as it melts. }
\label{Pu-phase}
\end{figure}              % End Figure

Plutonium is often described as ``A physicist's dream but an engineer's nightmare'' because the metal has an anomalously low melting point, a high number of allotropes, negative coefficient of thermal expansion in some phases, and the unusual preference for low-symmetry crystal structures. At ambient pressure Pu exhibits six solid allotropic crystal structures ($\alpha$, $\beta$, $\gamma$, $\delta$, $\delta$', $\epsilon$) between room temperature and melting, as shown in Fig~\ref{Pu-phase}. The phase stability of Pu as a function of temperature clearly illustrates the change in volume that occurs between the low-symmetry monoclinic structure of $\alpha$-Pu with 16 atoms in the unit cell and the high-symmetry structure of $\delta$-Pu. Indeed, despite a cubic structure, the fcc $\delta$-phase is perplexing given its unusually low density and negative thermal expansion coefficient, i.e., it contracts when heated. It is, however, most intriguing that the $\alpha$, $\beta$,  and $\gamma$ phases of Pu are low symmetry monoclinic or orthorhombic structures, a geometry that is common to minerals but foreign to metals and alloys which are usually cubic or hexagonal. An immediate question is why do the light actinides have low symmetry crystal structures if most all other metals have high symmetry atomic geometries? This is quite important given that the high symmetry structures of most metals, such as fcc copper and aluminium, exhibit ductility by readily allowing dislocation motion through 12 slip systems. It is the ductile behavior of metals that makes them prized for many technological applications. Accordingly, the low symmetry and concomitant brittle nature of Pu and other light actinides is unwanted and we must understand its physical origins.

\subsection{Lowering the Electronic Energy Through a Peierls-like Distortion}
\label{sec:lowering}

Protactinium, uranium, neptunium, and plutonium all exhibit low-symmetry ground-state structures rather than the high-symmetry structures usually found in metals. In the early days of actinide research, this low symmetry was attributed to directional or covalent-like bonding resulting from the angular characteristics of $f$ electrons. However, this so-called chemist view that the pointy $f$-electron atomic orbitals are forming directional bonds, hence no closed-packed structure of the metals, falls short of explaining the abundance of crystal structures observed of the actinide elements. Accordingly, physicists looked for a better answer and found it in the {\it{width}} of the band. S\"{o}derlind {\it{et al.} }\cite{soderlind95} showed that as the bandwidth of a bonding state is narrowed the crystal structure of the metal will distort through a Peierls-like distortion to a low-symmetry geometry. The original Peierls-distortion model was formulated in a one-dimensional lattice: A row of equidistant atoms can lower its total energy by forming pairs. The lower periodicity causes the degenerate energy levels to split into two bands with lower and higher energies. The electrons occupy the lower levels, so that the distortion increases the bonding and reduces the total energy. In one-dimensional systems, the distortion opens up an energy gap at the Fermi level making the system an insulator. However, in the higher dimensional systems, the material remains a metal after the distortion because other Bloch states are filling the gap. This mechanism is very effective if there are many degenerate levels near the Fermi level---that is, if the energy bands are narrow with a large density-of-states \cite{boring00}. Symmetry breaking of a crystal structure through a Peierls-like distortion due to narrow bandwidths (and concomitant high density-of-states near the Fermi level) is true for $s$, $p$, $d$, and $f$ states, meaning that even aluminium can be expanded enough that a low-symmetry body-centered tetragonal structure is observed \cite{soderlind95}.  Of course for metals with $s$, $p$, and $d$ bands actively bonding, huge theoretical tensions must be applied to narrow the bands enough to lower the crystal structure symmetry.  However, the $5f$ bands of Pa, U, Np, and Pu that are bonding are narrow enough to yield low symmetry structures at ambient conditions. Thus, the narrow width of the $5f$ band causes the low-symmetry structures of the ground state, without the need to invoke an explanation involving directional bonding.

Our discussion of Pu and energy-lowering lattice distortions means that the metal is directly on the border between between itinerant and localized $5f$ electron behavior. The metal shows a monoclinic ground state and an fcc high-temperature phase, resulting in $5f$ states that are balanced between a Peierls distortion to a low symmetry structure due to stronger bonding of the $f$ states and a high-symmetry crystal structure with strong electron correlations with weaker bonding of the $f$ states. The fact that $\delta$-Pu exhibits a high symmetry fcc structure even though it has active, albeit weaker, $5f$ electron bonding is somewhat perplexing.  Surely the degree of bonding is less than in $\alpha$-Pu given that the ground-state phase is monoclinic.  However, the exceedingly high electronic specific heat (Sommerfield coefficient) of 35-64 mJ K$^{-2}$ mol$^{-1}$ (Refs.~\cite{wick80,lashley03,javorsky06}) tells us that there is appreciable electron weight at the Fermi level in $\delta$-Pu, and this could only be achieved with the $5f$ states.  In order to understand this apparent contradiction, we can turn our attention to a rare-earth metal that has $4f$ states which can be changed from bonding to non-bonding through temperature or pressure, all the while retaining its high symmetry fcc structure.

\subsection{Comparison with Cerium}
\label{sec:cerium}

A material with similar properties as plutonium, and of comparable complexity, is cerium, the first element in the row of the lanthanides which is known for the ambivalent character of its $4f$ states. It exhibits multiple crystallographic phases that are strong functions of temperature, pressure, and chemistry \cite{koskenmaki78}. At ambient pressure, Ce metal exhibits four allotropic phases between absolute zero and its melting temperature at 1071 K, namely $\alpha$ (fcc), $\beta$ (dhcp), $\gamma$ (fcc), and $\delta$ (bcc). Magnetic, nonmagnetic and superconducting behavior is also observed. There is broad consensus that the unusual properties of Ce and its compounds originate essentially from the interplay of strong electronic correlations between the Ce $4f$ electrons and hybridization between $4f$- and conduction-electron states. Phenomena such as intermediate-valence, which points to non-integer occupation of the $4f$ shell, or heavy-fermion behavior characterized by an extremely large contribution of the electronic specific heat, are prominent observations on such systems. But perhaps one of the most intriguing phenomena is the $\gamma$-$\alpha$ phase transition in Ce metal \cite{koskenmaki78}. It occurs under pressure at room temperature or, at ambient pressure, on cooling to low temperature. Complexity of the underlying mechanisms is reflected in the collapse of the atomic volume by $\sim$17\% preserving the fcc lattice symmetry, and the loss of the magnetic moment ($\gamma$-Ce is paramagnetic and $\alpha$-Ce non-magnetic). The $\gamma$-$\alpha$ phase boundary ends at a critical point where the two phases become indistinguishable.

The anomalous behavior of Ce at the $\gamma$-$\alpha$ phase transition is commonly described by mechanisms involving either a Mott transition \cite{johansson74} or Kondo hybridization \cite{allen82,lavagna82}. However, recent first-principle calculations provide evidence for a combination of effects, involving strongly correlated $4f$ electrons in both phases and screening as important ingredients \cite{demedici05,mcmahan03,zwolfl01}. The transition to the $\alpha$-phase is accompanied by a sizable increase of the hybridization between $4f$- and valence-electron states. It leads to a partial delocalization of the $4f$ states, i.e., the ground state occupation number, $n_f$, that is almost equal to one in the $\gamma$-phase is reduced in the $\alpha$-phase. This has been directly derived from recent measurements of resonant inelastic x-ray scattering on elemental Ce under pressure \cite{rueff06}. The experiments permit to conclude that the changes of the electronic structure at the phase transition mainly result from band formation of $4f$ electrons that concurs with reduced electron correlation and increased Kondo screening. The experiments also highlight the importance of double occupancy of the $4f$ states in the ground state for understanding the effects of electron correlation in this element, as stated in Ref.~\cite{mcmahan03}. Considering lattice dynamics, phonon density of states of $\alpha$ and $\gamma$-phases in Ce$_{0.9}$Th$_{0.1}$ by Manley {\it{et al.} } show only a very small difference in the vibrational entropy of the two phases, where most of the transition entropy can be accounted for with the crystal field and changes in the ground-state spin fluctuations \cite{manley03}. 

In compounds, Ce shows $\gamma$-$\alpha$-phase-like character depending on the $4f$-valence band hybridization strength. Interestingly, in $\alpha$-like compounds, such as CeFe$_2$ or CeCo$_5$, the Ce atoms carry an ordered magnetic $4f$ moment \cite{brooks93} in contradistinction to elemental $\alpha$-phase Ce metal \cite{koskenmaki78}. An important concept in the theoretical description of these compounds is the hybridization between the $3d$ states of the transition metal and the delocalized Ce $4f$ states. Ce readily reacts with hydrogen. As with other light rare earth metals, it forms a cubic dihydride CeH$_2$ with hydrogen atoms on tetrahedrally coordinated sites (CaF$_2$-type structure) and dissolves, in a single phase, further hydrogen on octahedral sites up to the cubic trihydride CeH$_3$ (BiF$_3$-type structure) \cite{vajda95}. As the composition of the trihydride is approached a metal-to-insulator transition occurs that is reversible. It was recognized that thin films of such rare-earth hydrides allow to rapidly switch between the contrasting optical properties of the dihydride and trihydride phases, making them technologically interesting as optical switches \cite{huiberts96}. Model calculations of the rare earth hydrides in an ionic picture have shown the importance of electron correlations for this effect \cite{ng97,ng99,eder97}. Strong Coulomb interactions between the electrons on hydrogen sites were shown to be responsible for opening up a gap of $\sim$2 eV between the valence bands derived from rare earth-hydrogen and hydrogen-hydrogen hybridization and a set of bands predominantly of rare earth-metal $d$ character. X-ray absorption spectra have shown that the electronic configuration of the Ce $4f$ states in the Ce hydrides is similar as in the $\gamma$-phase of the pure metal \cite{arend99}. Below 7 K, CeH$_x$ orders magnetically in a complex phase diagram \cite{vajda95,arons91}.

\subsection{Stabilized $\delta$-Plutonium}
\label{sec:gallium}

Now back to the point of why we introduced Ce. $\alpha$-Ce has actively bonding $f$ states, but the crystal structure is fcc, as observed for $\delta$-Pu. For $\alpha$-Ce, this means there is just enough $f$ bonding to influence the physical properties such as electronic specific heat, but not enough to reduce the crystal structure. Of course this can easily be changed with external influences.  When pressurized to about 4 GPa, the structure changes to orthorhombic $Cmcm$ (the same as $\alpha$-U) meaning the $4f$ states become decisive enough in bonding so as to break the cubic symmetry of the crystal.  Considering $\delta$-Pu, a similar situation is occurring where there is enough $f$ bonding to influence the electronic specific heat, but not enough to break cubic symmetry. It is possible there is a slight distortion from a perfect cubic symmetry in $\delta$-Pu, but if true this would be small \cite{valone06,moore06b}. Once $\delta$-Pu is pressurized to about 0.1 GPa, the structure looses cubic symmetry. In a reverse manner, adding a few \% of Ga, Al, Ce, Am, or other tetravalent elements stabilizes $\delta$-Pu at room temperature and below. 

The addition of impurity atoms often destroys the coherence of the $f$ band. Without its narrow $f$ band actively bonding, plutonium can no longer reduce its energy by lowering its symmetry through Peierls-like distortion to the $\alpha$-phase; it therefore remains in the $\delta$-phase. Another view is that plutonium atoms relax and move toward the smaller non-$f$ atoms, thereby reducing the $f$-$f$ interactions that stabilize the $\alpha$-phase. Either way, we see that addition of many different elements stabilizes $\delta$-Pu, such as Al, Ce, Am, and, most often utilized, Ga. Indeed, we find $\delta$-phase stabilization occurs even for defects, where it has been found that excessive plastic deformation, and, concomitantly the introduction of dislocations, stabilizes $\delta$-Pu \cite{gorbunov01}. Once more we see how external influences, such as temperature and chemistry as well as the pressure or defects can influence the phase stability of plutonium. What is more, we now have a way to overcome the crippling technological disadvantage of monoclinic Pu that is brittle and difficult to machine.

\section{Revised View of the Periodic Table}
\label{sec:perodic-table}

\begin{figure}[t]           %FIGURE PTable           FIG. 3
%\sidecaption[t]
\begin{center}
\includegraphics[scale=0.7]{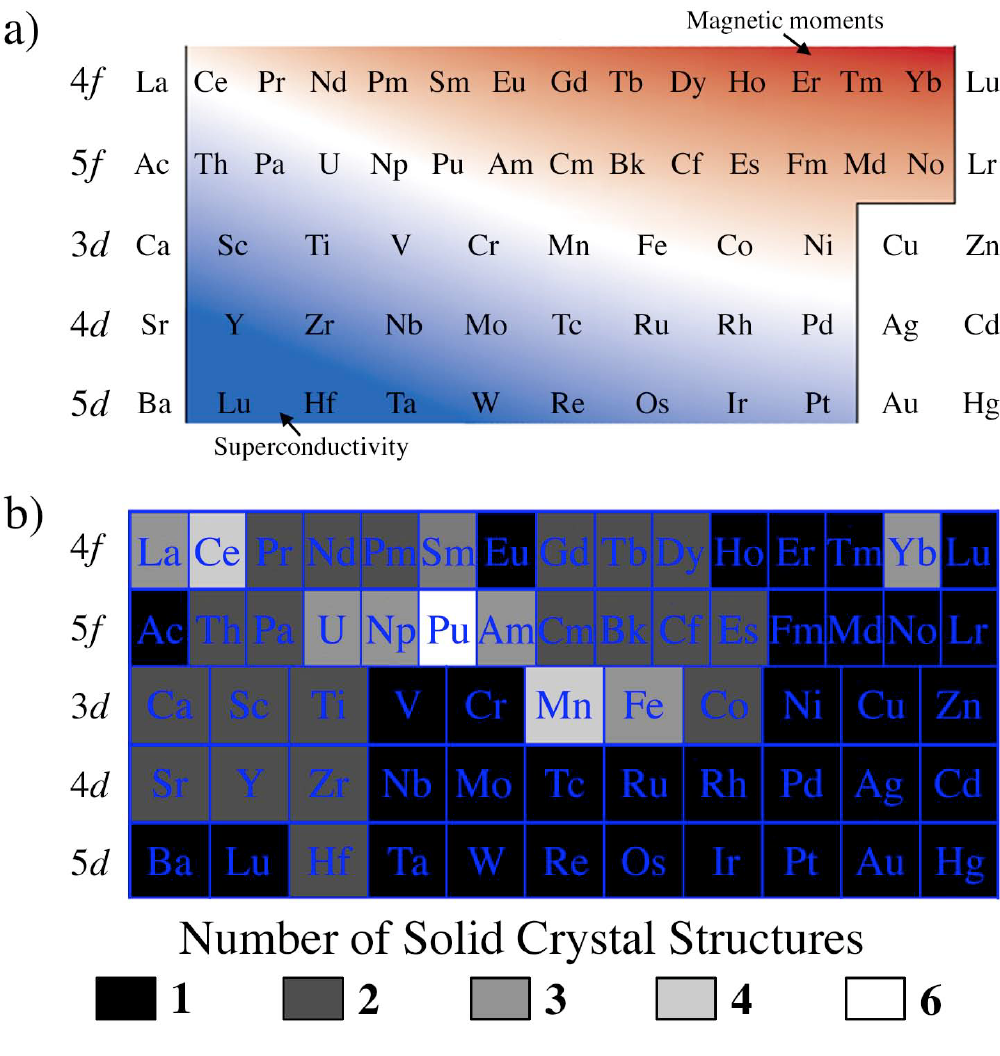}
\caption{ 
(Color online) Rearranged Periodic Table showing the five transition metal series, $4f$, $5f$, $3d$, $4d$, and $5d$ (after Refs.~\cite{smith83} and \cite{boring00}).  When cooled to the ground state, the metals in the blue area exhibit superconductivity while the metals in the red area exhibit magnetic moments.  The white band running diagonally from upper left to lower right is where conduction electrons change from itinerant and pairing to localized and magnetic.  Slight changes in temperature, pressure, or chemistry will move metals located on the white band to either more conductive or more magnetic behavior. (b) Version of (a) where the number of solid allotropic crystal structures for each metal is indicated by gray scale.  Lighter shades indicate more phases.  Notice that a band of lighter shades mirrors the white band in (a), showing that metals on or near the transition between magnetic and superconductive behavior exhibit numerous crystal phases.  }
\label{PTable}
\end{center}
\end{figure}            % End Figure

Plutonium is not alone in juggling its electrons between bonding and localized states. Figure~\ref{PTable} shows a rearranged Periodic Table containing the $d$ and $f$ electron series, where the $f$ series are placed on top and spaced slightly tighter. The localization increases going from bottom to top. Furthermore, along each series the localization increases from left to right. The elements on the white diagonal stripe form the rough dividing line between localized (local moment magnetism) and itinerant (superconductivity) long-range collective behavior. The elements on or near the transition between localized and delocalized behavior, i.e., the white diagonal stripe, exhibit a large number of solid allotropic phases [see Fig.~\ref{PTable}(b)]. This is clearly illustrated in the lower part of Fig.~\ref{PTable} where the lighter shades indicate more phases, and the light band mirrors the white diagonal stripe in the rearranged Periodic Table above \cite{moore04a}. The most notable elements with numerous phases are: Pu with 6, Ce with 4, Mn with 4, as well as La, U, Np, Am, and Fe, each with 3.

The elements at the diagonal white stripe in Fig.~\ref{PTable} have in common that the wave functions from different atoms are barely overlapping, and their electrons are thus bordering on being localized. These are the so-called correlated-electron materials that are characterized by the presence of a narrow conduction band, giving rise to spin and charge fluctuations associated with the low-energy excitations. The exotic behaviors of these materials necessitate a description by many-electron models, such as the Kondo, Hubbard, or Anderson models. These models can be generically classed as two-electron ``impurity'' models, which introduce interactions between pairs of electrons, one localized on an impurity atom and one in a conduction band. The Anderson Hamiltonian contains both a repulsive Coulomb term and an electron hopping term. The former term kepdf the $5f$ electrons localized, and the latter term leads to a partial localization of the conduction electrons. \cite{boring00}

\section{Actinide Magnetism}
\label{sec:actinide-magnetism}
%\subsection{Localization and Magnetic Moment Formation}

The relation between $f$-electron localization and the formation of magnetic moments, as illustrated in Fig.~\ref{PTable}(a), leads to the following conclusions. The light actinides (Th, U, Np) are delocalized and show no magnetic ordering. Many of these metals are even superconducting at low temperatures. On the other hand, the localized heavy actinides (from Cm onwards) show magnetic ordering. Note that americium is $5f^6$ with a ground state $J=0$ and therefore has no magnetic ordering. Thus, magnetism---or lack thereof---of most of the actinides is clear. The situation regarding the magnetism of Pu, however, remains murky.

\subsection{Experimental Absence of Magnetic Moments in Plutonium}
\label{sec:experimental-absence}

Plutonium shows a large experimentally observed magnetic susceptibility, hints of heavy-fermion behavior, and an anomalous temperature dependence of the electrical resistivity.  At low temperatures the resistivity increases, which is characteristic for a Kondo lattice. Judging by its small volume the monoclinic $\alpha$-phase is clearly delocalized and therefore non-magnetic. The fcc $\delta$-phase is more localized. So, could this phase be magnetic? In the past, {\it{ab-initio}} electronic structure calculations that included spin and orbital polarization successfully reproduced the large volume change by localizing the $5f$ electrons in $\delta$-Pu \cite{soderlind94, Antropov95, Savrasov00, Kutepov03}. One particular set of calculations by S\"oderlind and Sadigh \cite{soderlind04} achieved the appropriately spaced energies and atomic volumes for five of the six allotropic phases of Pu (the high-temperature bcc $\epsilon$-phase was too high in energy) as well as an equation-of-state, bulk modulus, and elastic constants that are in agreement for most all of the six allotropic phases. However, these calculations gave rise to a new problem, namely they concluded that Pu contained local magnetic moments. While experimental results in the early days were ambiguous, today it well established that all six phases of Pu, including $\delta$, show no evidence of long-range magnetism.

\begin{figure}[t]           %FIGURE    Suscep           FIG. 4
\sidecaption[t]
\includegraphics[scale=0.45]{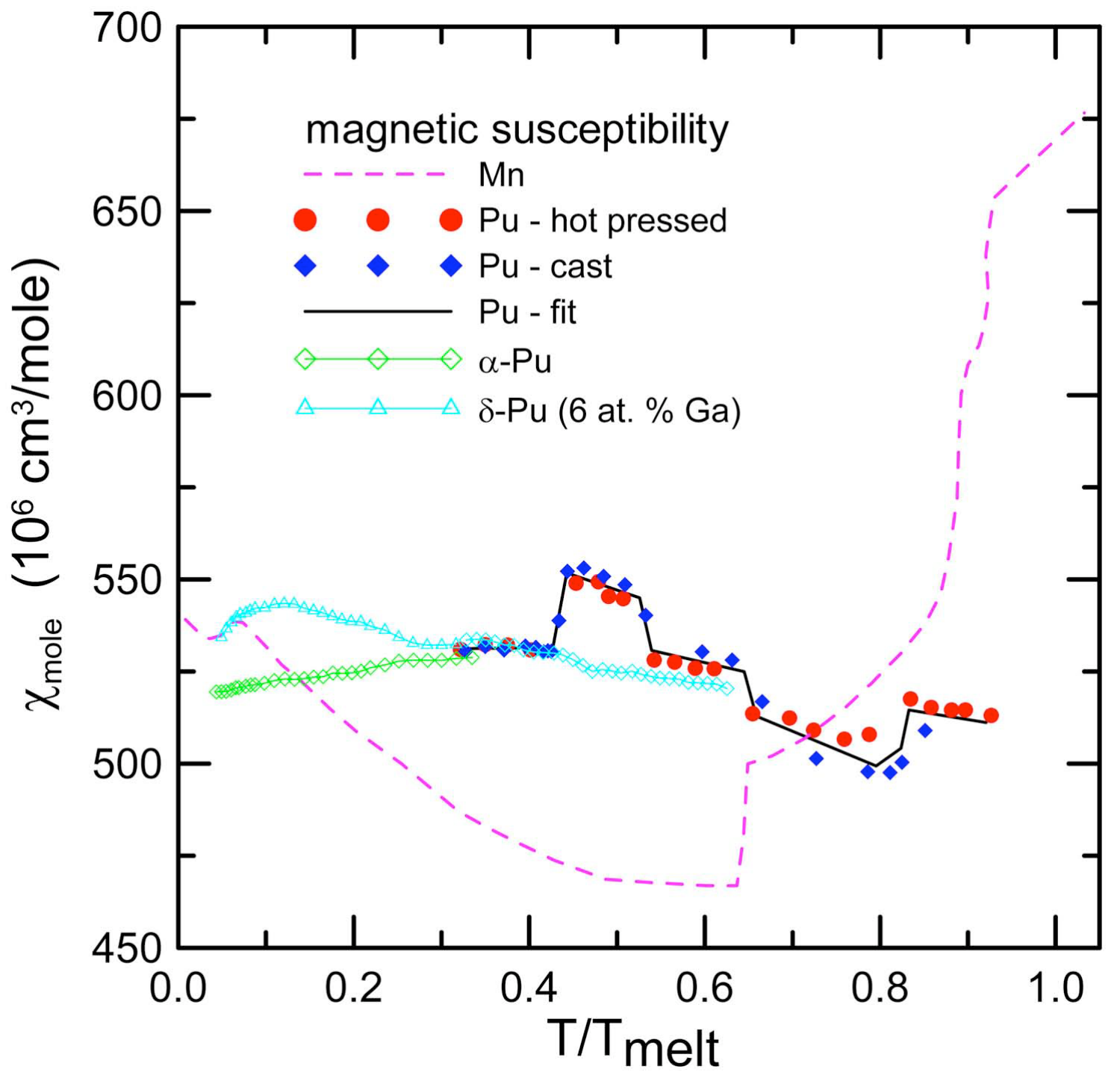}
\caption{ 
Molar susceptibility of Mn and various forms of Pu plotted as a function of the melting point [$T_{\mathrm{melt}}$(Mn) =1519 K and ($T_{\mathrm{melt}}$(Pu) = 913 K].    The susceptibility of Pu is characteristic of metals with relatively strong paramagnetism.
Manganese is given for comparison. The sequence for Mn is $\alpha$ (complex cubic), $\beta$ (complex cubic), $\gamma$ (fcc), and $\delta$ (bcc). Mn orders antiferromagnetically at 95 K ($T/T_{\mathrm{melt}}$ = 0.063) in a complex tetragonal structure that is a slight distrotion of the $\alpha$-phase. From Ref.~\cite{lashley05}. 
 }
\label{Suscep}
\end{figure}            % End Figure

Magnetic measurements on Pu were reviewed in 2005 by Lashley {\it{et al.}} \cite{lashley05} In order to understand the large volume expansion between $\alpha$- and $\delta$-phases of plutonium a {\it{localization}} of the $5f$ states is needed. With strong reasons for taking the number of $5f$ electrons in Pu as five, this gives a Kramer's ground state with the prediction of magnetic ordering for the $\delta$-phase, and in some cases, even for the $\alpha$-phase. The resulting moments are predicted to be large, and even though, due to their antiparallel coupling there is a partial cancellation of the spin and orbital parts, the total magnetic moments are of the order of 1-2 $\mu_{\mathrm{B}}$/atom.  This remains the case when band structure calculations take into account orbital polarization effects, in which case the absolute value of the orbital magnetic moment exceeds the value of the spin magnetic moment. 

The measured magnetic susceptibilities of Pu in its various phases is shown in Fig.~\ref{Suscep} (Ref.~\cite{lashley05}. This susceptibility is characteristic of metals with relatively strong paramagnetism caused by electronic band magnetism. Neither the temperature nor magnetic-field dependences of the measured susceptibilities in Pu provide evidence for disordered or ordered moments. Specific-heat measurements indicate the absence of magnetic entropy. The combination of neutron elastic and inelastic scattering data shows no convincing evidence for either long-range ordered or disordered (static or dynamic) magnetic moments. On the basis of resistivity measurements for both $\alpha$- and $\delta$-Pu, and review of other measurements, Arko, Brodsky, and Nellis \cite{arko72} concluded in 1972 that there was {\it{no evidence for localized moments or magnetic order}}. On the basis of a $T^2$ dependence of the resistivity near $T$= 0 these authors suggested a model involving spin fluctuations for elemental Pu, as well as for a number of other alloys and compounds.

Review of the magnetic measurements by Lashley {\it{et al.}} \cite{lashley05} demonstrated beyond all reasonable doubt that there is no {\it{ordered}} magnetism involving the $5f$ electrons in Pu metal in either the $\alpha$- or $\delta$-phases down to a base temperature of $\sim$4 K. The experimental evidence presented includes magnetic susceptibility, specific heat (with an applied field of up to 14 T), nuclear magnetic resonance, and elastic and inelastic neutron scattering. Previous reports of ``anomalies'', seen especially in the specific heat, can in fact be ascribed to impurities or structural effects, most probably the occurrence of martensitic transformations of some parts of the samples to the $\alpha$'-phase that contains Ga. The absence of any diffuse scattering in the neutron diffraction patterns (except at high-temperature from thermal disorder) also argues against any disordered local moments. Dynamical mean-field theory (DMFT) results by Savrasov, Kotliar, and Abrahams \cite{savrasov01} imply, that the local moments are "washed out" over short time scales and thus may not be observable to probes such as NMR and neutron inelastic scattering, depending on the probes' observational frequency window. Again, more effort to quantify these predictions for experiments on Pu would seem worthwhile. \cite{lashley05,heffner06} 

\subsection{Looking to other elements for clues}

As plutonium has no local magnetic moments, a limited comparison could perhaps be drawn with palladium, which finds itself also on the diagonal stripe in Fig.~\ref{PTable}(a). As bulk compound, palladium is an anomalous paramagnet on the verge of ferromagnetism \cite{vitos00,staunton00}. The free Pd atom has a [Kr]$4d^{10}$ closed-shell electronic configuration. In bulk Pd no spontaneous ferromagnetic order is observed. Although the Fermi level is located immediately above a sharp peak in the density of states, at the equilibrium volume the Stoner criterion is not satisfied. The most sensitive factors that may enhance the density of states at the Fermi level, and cause the onset of ferromagnetism in Pd clusters are the electron localization associated with a reduced coordination number, the expansion of the lattice, and the change of the local symmetry \cite{vitos00}.

One might also ask whether the conduction electrons of Pu could make it an itinerant magnet, like iron. While there is little evidence for magnetism in pure plutonium, many plutonium compounds are indeed magnetic and tend to be itinerant magnets \cite{santini99}. Indeed, simply dissolving hydrogen in plutonium is enough to make the electrons localize and to turn the system ferromagnetic. Also, a comparison of the light actinides with the transition metals indicates that the light actinides should be superconductors unless they have local moments. So, the fact that plutonium is not a superconductor might indicate that plutonium is an incipient, weak itinerant magnet and that the loss of magnetic ordering with heating plays a role in the contraction of the $\delta$-phase.

\section{Experimental Complications of Plutonium}
\label{sec:experimental-complications}

Due to its toxic and radioactive nature, hands-on investigation of Pu can only be done at a small number of institutions around the world, making experimental results sparse in comparison to other materials. Furthermore, experiments on Pu suffer from the problem of self-irradiation, which heats the sample, slowly destroys the crystal structure \cite{schwartz05,moore07c}, grows in daughter products of the decay process, and potentially creates local magnetic moments \cite{mccall06}. When a plutonium nucleus in the material undergoes $\alpha$-decay, the recoiling uranium and helium nuclei knock plutonium atoms from their lattice sites. Displaced plutonium atoms come to rest at interstitial sites and leaving lattice vacancies. Each displaced plutonium atom creates a Frenkel pair---consisting of a vacancy and a self-interstitial. Each decay event creates more than 2000 Frenkel pairs. The self-heating means that in practice it is extremely difficult to achieve temperatures much below $\sim$2 K, unless very small ($<$mg) samples are used, and even then there is always some doubt as to the real temperature of the sample.  This, however, may be overcome by choosing the right isotope. The most common isotope is fissile $^{239}$Pu with a 24,400 year half-life, which is made in nuclear reactors: $^{238}$U + neutron $\to$ $^{239}$Pu. The 239-isotope produces considerable self heating.  Alternatively, $^{242}$Pu has a 376,000 year half-life, which drastically reduces heat due to self irradiation.  Finally, the cr\`eme de la cr\`eme is $^{244}$Pu with a $\sim$80 million year half-life, meaning the isotope is near non-radioactive.  This isotope, if separated in usable quantities (it is very rare), could open many new avenues for low-temperature experiments on Pu.

\section{One Man's Electron Energy Loss  is Another's X-Ray Absorption}
\label{sec:substituting}

While the technique of x-ray absorption spectroscopy has become routine and is the go-to measurement for physicists, actinides and their special requirements demand a different experiment: Electron energy-loss spectroscopy (EELS) in a transmission electron microscope (TEM). Why is this? First, the TEM utilizes small samples, allowing one to avoid the handling of appreciable amounts of toxic and radioactive materials. The alternative XAS performed at a multi-user synchrotron radiation facility is usually less well-adapted for the delicate and secure handling of radioactive materials. Second, the technique is bulk sensitive due to the fact that the electrons traverse $\sim$40 nm of metal, this being the appropriate thickness for quality EELS spectra of actinide materials at the primary energy used in our TEM. A few nanometers of oxide do form on the surfaces of the TEM samples, but this is insignificant in comparison to the amount of metal sampled through transmission of the electron beam. Third, actinide metals near the localized-itinerant transition exhibit numerous crystal structures that can coexist in metastable equilibrium due to the small energy difference between the phases \cite{moore04a}. Therefore, acquiring single-phase samples of metals at or near this transition is uncertain, making spectroscopic techniques with low spatial resolution questionable. Finally, actinide metals readily react with hydrogen and oxygen, producing many unwanted phases in the material during storage or preparation for experiments.  The TEM has the spatial resolution to image and identify secondary phases \cite{hirsch77,reimer97,fultz01}, ensuring examination of only the phase(s) of interest. A field-emission-gun TEM, such as the one used in these experiments, can produce an electron probe of $\sim$5 \AA\, meaning recording spectra from a single phase when performing experiments is easily achieved. Quantitatively measuring the reflections in the electron diffraction pattern, as well as other crystallographic orientations, proves that the correct phase is examined \cite{zuo92,moore02}. Hence there is no need for large single crystals!

EELS spectra of the actinide metals have been collected at the Lawrence Livermore National Laboratory (LLNL) using a Phillips CM300 field-emission-gun TEM, equipped with a Gatan imaging filter \cite{moore04a}. The momentum transfer in EELS gives rise to multipole transitions. However, by employing a small objective aperture and using a high primary energy of the electron beam (297 kV) the low-energy transitions are electric-dipole. The similarity between high-energy EELS spectra and synchrotron-radiation-based XAS has been well established for some time, and was recently validated for $f$-electron systems for the case of Ce metal \cite{moore04b}. 

\section{Theory}
\label{sec:theory}

The behavior of the $5f$ electrons in the actinides is governed by the interplay of the spin-orbit  and electrostatic interactions. 
Here, we treat the effect of these interactions on the electronic configuration in the different angular momentum coupling schemes, i.e., $jj$-, $LS$-, and intermediate coupling. 
  This will be illustrated for the example of the  $f^2$ configuration. This two-particle state is rather straightforward as it does not require explicit use of coefficients of fractional parentage or creation and annihilation operators. It is also shown how to obtain the expectation values of the moments in each of the different coupling schemes. For further reading on these topics we refer  e.g. to the {\emph{Hitchhiker's Guide to Multiplet Calculations}}
 \cite{vanderlaan06}.

\subsection{Atomic Interactions}
\label{sec:el-and-so}

For $n$ electrons moving about a point nucleus of charge, the Hamiltonian can be written in the central field approximation as
\begin{equation}         % EQUATION 1
{\vec{H}} = {\vec{H}}_{\mathrm{el}} + {\vec{H}}_{\mathrm{so}} \; ,
\label{eq:1}
\end{equation}				
\noindent
where ${\vec{H}}_{\mathrm{el}}$ and ${\vec{H}}_{\mathrm{so}}$ are the terms for the electrostatic and spin-orbit interaction, respectively.
This atomic Hamiltonian can be embedded in a solid state, e.g., using an Anderson impurity model \cite{vanderlaan00}.
Crystal field interactions are usually less prominent in the actinides, leading only to a small perturbation, therefore we assume a spherical potential. In any case, all interactions can be separated in {\it{angular}} and {\it{radial}} parts. The angular parts depend on the angular quantum numbers of the basis states of the electronic configuration and are independent of the radial wave functions, which  in the calculations are taken as empirical scaled parameters (see Sec.~\ref{sec:IC}). 

The basis wave functions are assumed to be an antisymmetrized product of one-electron functions. In spherical symmetry, these basis states are eigenfunctions of the total angular momentum $J$ and its component $M_J$. The states are characterized by quantum numbers $\alpha LS$, where $\alpha$ is a suitable quantity for distinguishing between terms having the same values of the orbital and spin angular momenta, $L$ and $S$.

\subsubsection{Electrostatic Interactions}
\label{electrostatic}

The non-relativistic Hamiltonian for the electrostatic interactions of $n$ electrons in an atom with nuclear charge $Ze$  is 
\begin{equation}      % EQUATION 8
{\vec{H}}_{\mathrm{el}}= -\frac{\hbar^2}{2m} \sum_{i=1}^n \nabla^2_i - \sum_{i=1}^n \frac{Ze^2}{r_i} + \sum_{i<j}^n \frac{e^2}{r_{ij}} \;.
\label{eq:8}
\end{equation}

The first term describes the kinetic energy of all electrons, the second one gives the potential energy of all electrons in the potential of the nucleus. The third term describes the repulsive Coulomb potential of the electron-electron interaction, where the expectation value can be expressed as
\begin{equation}      % EQUATION 9
\left\langle \alpha LS \left| \frac{e^2}{|r_1-r_2|} \right| \alpha LS  \right\rangle = \sum_k f_k F^k +  \sum_k g_k G^k \;,
\label{eq:9}
\end{equation}

\noindent
with angular parts $f_k$ and $g_k$ and radial parts $F^k$ and $G^k$. 

The radial integrals $F^k$ and $G^k$ are theoretically defined by the Slater integrals \cite{condon63} and experimentally treated as empirically adjustable quantities to fit the observed energy levels and their intensities.
The direct integrals $F^k$ represent the actual electrostatic interaction between the two electronic densities of electrons $ \ell$ and $ \ell '$. The exchange integrals $G^k$ arise due to the quantum mechanical principle that fermions are indistinguishable, so that the wave function is totally antisymmetric with respect to permutation of the particles. Consequently, $G^k$ is not present  for the configuration $\ell^n$, where the electrons are equivalent. 

The angular parts $f_k$ and $g_k$ can be written in terms of 3$j$- and 6$j$-symbols \cite{cowan81}
\begin{eqnarray}      % EQUATIONS 9a and 9b
f_k (\ell, \ell') &=& (-1)^L  [\ell,\ell'] 
\left( \begin{array}{ccc} \ell & k & \ell \\ 0 & 0 & 0 \end{array} \right)
\left( \begin{array}{ccc} \ell' & k & \ell' \\ 0 & 0 & 0 \end{array} \right)
\left\{ \begin{array}{ccc} \ell & \ell'  & L \\   \ell' & \ell & k \end{array} \right\} \; , 
\label{eq:9a} \\
g_k (\ell, \ell') &=& (-1)^S  [\ell,\ell'] 
\left( \begin{array}{ccc} \ell & k & \ell' \\ 0 & 0 & 0 \end{array} \right)^2
\left\{ \begin{array}{ccc} \ell & \ell'  & L \\   \ell & \ell' & k \end{array} \right\} \; .
\label{eq:9b} 
\end{eqnarray}

\noindent
The triangle conditions require that $f_k(\ell,\ell)$ has non-zero values for $k=0,2, \dots, 2\ell$, whereas $g_k(\ell,\ell')$ has non-zero values for $k=|\ell-\ell'|, |\ell-\ell'+2| \dots, \ell+\ell'$, with $\ell \ne \ell'$.  Thus the initial state $f^n$ has the radial parameters $F^0$, $F^2$, $F^4$, $F^6$, while the final state $d^9f^n$ has $F^2$, $F^4$, $G^1$, $G^3$, $G^5$.

\subsubsection{Spin-Orbit Interaction}
\label{sec:so}

Turning to the second term of Eq.~(\ref{eq:1}), the spin-orbit interaction for the $\ell$ shell is given as 
\begin{equation}      % EQUATION 2
{\vec{H}}_{\mathrm{so}} = \zeta_{\ell}(r) \sum^n_{i=1} {\vec{l}}_i \cdot {\vec{s}}_i  \; ,
\label{eq:2}
\end{equation}

\noindent
where ${\vec{l}}_i $ and ${\vec{s}}_i$ are the one-electron orbital and spin angular momentum operators of the $i$-th electron of the $\ell^n$  configuration.  $\zeta_{\ell}(r)$ gives the radial part and $\sum_{i} {\vec{l}}_i \cdot {\vec{s}}_i$ gives the angular part of the spin-orbit operator. For brevity, we also introduce the following notation                      
\begin{equation}       % EQUATION 3
{\vec{l}}\cdot {\vec{s}}  =  \ell s \, w^{110}
  \equiv  \sum^n_{i=1} {\vec{l}}_i \cdot {\vec{s}}_i \; ,
\label{eq:3}
\end{equation}

\noindent
where $w^{110}$ is the coupled tensor for the spin-orbit interaction (see e.g. Refs.~\cite{vanderlaan96,vanderlaan98} for details on coupled tensors).
The expectation value for a one-electron state $| \ell s j \rangle$  with $\ell \neq 0$ is 
\begin{equation}        % EQUATION 4
\langle \ell s j | {\vec{l}} \cdot  {\vec{s}} | \ell s j \rangle
=\frac{1}{2} \left[ j(j+1)-\ell(\ell+1)-s(s+1) \right] \; ,
\label{eq:4}
\end{equation}

\noindent
which gives a doublet $|j=\ell \pm s \rangle$ with expectation values
\begin{equation}      % EQUATION 6
\langle \ell s j | {\vec{l}} \cdot  {\vec{s}} | \ell s j \rangle 
= \left\{ \begin{array}{ccc} 
-\frac{1}{2}(\ell+1) & {\mathrm{for}} & j_1=\ell-s \;,\\
\frac{1}{2} \ell &{\mathrm{for}} & j_2=\ell+s  \;.
\end{array}  \right .
\label{eq:6}
\end{equation}

For the atomic many-electron state, the Hamiltonian $\vec{H}_{\mathrm{so}}$ commutates with ${\mathbf{J}}^2$ and $J_z$ and is hence diagonal in $J$ and independent of the magnetic quantum number $M_J$. However, it does not commute with ${\mathbf{L}}^2$  or ${\mathbf{S}}^2$ and can thus couple states of different $LS$ quantum numbers.
For a pure state $\left| \ell^n \alpha LSJ \right\rangle $ the expectation value is
\begin{equation}         % EQUATION 5
\langle \ell^n \alpha LSJ | {\vec{l}} \cdot {\vec{s}} | \ell^n \alpha LSJ \rangle = \frac{1}{2}   [J(J+1)-L(L+1)-S(S+1)] \frac{\zeta(\alpha LS)} {\zeta} \; ,
\label{eq:5}
\end{equation}

\noindent
where the factor $ \zeta(\alpha LS)/ \zeta $ is equal to $n^{-1}$, 0, and $-(4\ell+2-n)^{-1}$ for $n <  2 \ell +1 $, $n = 2\ell+1$, and $n>2\ell+1$, respectively.

It is useful to have a general expression for $\langle {\vec{l}} \cdot {\vec{s}} \rangle$  valid for any many-electron state in intermediate coupling, including the $LS$- and $jj$-coupling limits.  
Since the spin-orbit operator is diagonal in $j$, there are no cross terms between $| j_1 \rangle$ and $| j_2 \rangle$. Therefore, we can distribute $n$ over $n_{j_1}$ and $n_{j_2}$, which are the  electron occupation numbers of $| j_1 \rangle$ and $| j_2 \rangle$. 
For the configuration $\ell^n$ with $n=n_{j_1}+n_{j_2}$,
application of Eq.~(\ref{eq:4}) gives the anticipated general expression
\begin{equation}    % EQUATION 7
\langle \ell^n J | \mathbf{l} \cdot \mathbf{s} | \ell^n J \rangle =
\sum_{j=j_1,j_2} \langle j | \mathbf{l} \cdot \mathbf{s} | j \rangle n_j =
-\frac{1}{2}(\ell+1)n_{j_1} +\frac{1}{2} \ell n_{j_2 } \; ,
\label{eq:7}
\end{equation}

\noindent
which does not depend on the values of $L$, $S$, and $J$.

\subsection{$LS$- and $jj$-Coupling Schemes}

The electrostatic interaction is diagonal in the $LS$-coupled basis and the spin-orbit interaction is diagonal  in the $jj$-coupled basis. Switching between the $LS$- and $jj$-coupled basis states is done using a transformation matrix containing the recoupling coefficients.

\subsubsection{$LS$ Coupling}

In the $LS$-coupling scheme, which gives the eigenfunctions in the limit  $\zeta_{\ell}(r) \to 0$, the particles are coupled as
\begin{equation}    % EQUATION 10
[(\ell_a,\ell_b)L,(s_a,s_b)S]J \;.  
\label{eq:10}
\end{equation}

\noindent
The Russell-Saunders notation for the $LSJ$-coupled states is $^{2S+1}L_J$.
The lowest energy term of a configuration $\ell^n$, i.e., the Hund's rule ground state, is that term of maximum $S$ which has the largest value of $L$. In addition, according to the third Hund's rule, $J=L-S$ ($J=L+S$) for less (more) than half filled shell.

\subsubsection{$jj$ Coupling}

In the $jj$-coupling scheme, which gives the eigenfunctions  in the limit  $\{F^k,G^k\} \to 0$, the particles are coupled as 
\begin{equation}      % EQUATION 11
[(\ell_a, s_a)j_a,(\ell_b,s_b)j_b]J \;.  
\label{eq:11}
\end{equation}

\noindent
In the $jj$-coupled ground state, first all $j=\ell-s$ levels are filled prior to the $j=\ell+s$ levels. 
For equivalent electrons, i.e., with equal values of $j$, the Pauli exclusion principle limitations on the possible values of $m_j$ prohibit some of the $J$-values that would be predicted by a vector model \cite{cowan81}. The result is that for any two equivalent electrons $\ell^2$ only the even $J$-values are allowed when $j_a = j_b$. 

\subsubsection{Transformation Matrix}

For a given electronic configuration, we arrive in both coupling schemes at the same set of allowed values of the total angular momentum $J = |L - S|, \dots , L + S $. This means that the Hamiltonian of Eq.~(\ref{eq:1}) is block-diagonal in $J$. For each $J$-block the basis states can be transformed between $LS$- and $jj$-coupling using recoupling coefficients that can be expressed in terms of $9j$ symbols,
\begin{equation}         % EQUATION 12
\left\langle  [(\ell_a,\ell_b)L,(s_a,s_b)S]J   |   [(\ell_a, s_a)j_a,(\ell_b,s_b)j_b]J  \right\rangle 
= [ L,S,j_a,j_b]^{\frac{1}{2}} \left\{ \begin{array}{ccc}
\ell_a & \ell_b & L \\
s_a & s_b & S \\
j_a & j_b & J
\end{array} \right\} \;,
\label{eq:12}
\end{equation}

\noindent
where $[x,y,\dots] \equiv (2x+1)(2y+1) \dots$.
These coefficients  form a matrix ${\vec{T}} = \langle  LS  \mid  jj \rangle$.
For a given $J$-block,  transformation of the  wave function or operator
is given by
\begin{equation}        % EQUATION 13
\begin{array}{lll}
| jj \rangle = {\vec{T}}  | LS \rangle & \,\, \mathrm{and} \,\, &  |LS  \rangle = {\vec{T}}^{-1} | (jj) \rangle \;, \\
H^{(jj)} = {\vec{T}} \, {\vec{H}}^{(LS)}  \, {\vec{T}} ^{-1} & \,\, \mathrm{and} \,\, & 
{\vec{H}}^{(LS)} = {\vec{T}}^{-1} {\vec{H}}^{(jj)}   {\vec{T}} \; .
\end{array}
\label{eq:13}
\end{equation}

\subsubsection{$LS$- Versus $jj$-Coupled Ground State: Example for $f^2$}
\label{sec:Ls-vs-jj}

To make the angular momenta coupling more explicit, let us examine the two-electron case, $f^2$. The possible $LSJ$ states for this configuration are $^1S_0$, $^1D_2$, $^1G_4$, $^1I_6$, $^3P_{0,1,2}$, $^3F_{2,3,4}$, $^3H_{4,5,6}$, amounting to a total of $\sum_i (2L_i+1)(2S_i+1)$ =  $\sum_i (2J_i+1)$ = 91 $M_J$ levels.
For  two equivalent $f$ electrons, the Pauli principle allows the $jj$ states (5/2,5/2)$_{0,2,4}$, (5/2,7/2)$_{1,2,3,4,5,6}$,   (7/2,7/2)$_{0,2,4,6}$. This also gives a total of $\sum_i (2J_i+1)$ = 91 $M_J$ levels, which illustrates that the Hamiltonian in Eq.~(\ref{eq:1}) is block diagonal in $J$.

In any coupling, including intermediate, the ground state has  total angular momentum $J$ = 4. The $LS$-coupled Hund's rule ground state is $|^3H_4 \rangle$ 
and the $jj$-coupled ground state is $|(5/2,5/2)_4 \rangle$.
Spin-orbit interaction  mixes  $|^3H_4 \rangle$   with $|^1G_4\rangle$ and $|^3F_4\rangle$.  The transformation matrix for $( | ^3H_4 \rangle,  | ^1G_4  \rangle,  |^3F_4  \rangle )$ 
$\to$ $(|(5/2,5/2)_4 \rangle, |(5/2,7/2)_4 \rangle, |(7/2,7/2)_4  \rangle )$
is obtained from Eq.~(\ref{eq:12}) as
\begin{equation}       % EQUATION 14
{\vec{T}}= \frac{1}{7} \left(
\begin{array}{ccc}
\sqrt{ \frac{110}{3} }      &   \sqrt{11}     & -\frac{2}{\sqrt{3}}  \\
 4 \sqrt{ \frac{2}{3} }      &    -2 \sqrt{5}  &  \sqrt{ \frac{55}{3} }  \\
-\sqrt{ \frac{5}{3} }         &    3 \sqrt{2}   &   2\sqrt{ \frac{22}{3} } 
\end{array}  \right) \; .
\label{eq:14}
\end{equation}

\noindent
Note that each state is ($2J+1$ = 9)-fold degenerate in the absence of a magnetic field. Using  Eq.~(\ref{eq:13}), the $jj$-coupled ground state  can be written as a sum over $LS$-coupled basis states
\begin{equation}   % EQUATION 15
|(5/2,5/2)_4 \rangle= \frac{1}{7} \sqrt{ \frac{110}{3}} \, | ^3H_4 \rangle + \frac{1}{7} \sqrt{11}\, |^1G_4  \rangle - \frac{2}{7\sqrt{3}} \,|^3F_4 \rangle \;.
\label{eq:15}
\end{equation}

\noindent
The character is given as the square of the wave function coefficient, so that the $jj$-coupled ground state has 74.8\% $|^3H_4\rangle$, 22.5\%  $|^1G_4 \rangle$, and 2.7\% $|^3F_4  \rangle$. This shows the $jj$-coupled ground state contains a considerable amount of low spin.

Equation~(\ref{eq:13})  also allows us to express the $LS$-coupled Hund's rule ground state as a sum over  $jj$-coupled basis states
\begin{equation}   % EQUATION 16
|^3H_4  \rangle = \frac{1}{7} \sqrt{ \frac{110}{3} } \, | (5/2,5/2)_4 \rangle + \frac{4}{7} \sqrt{ \frac{2}{3} } \,|(5/2,7/2)_4  \rangle - \frac{1}{7} \sqrt{\frac{5}{3} } \,| (7/2,7/2)_4  \rangle \;. 
\label{eq:16}
\end{equation}

\noindent
Thus the $jj$-coupled ground-state has 74.8\% $|(5/2,5/2)_4 \rangle$, 21.8\% $|(5/2,7/2)_4  \rangle$, and 3.4\% $|(7/2,7/2)_4  \rangle$ character. Since $ \langle  ^3H_4  \mid (5/2,5/2)_4 \rangle$ = 0.748, there is $\sim$75\% of the total electronic state  can be found simultaneously in  $|^3H_4  \rangle$ and $|(5/2,5/2)_4 \rangle$. The remaining $\sim$25\% is distributed over different states depending on whether the spin-orbit or Coulomb interaction prevails.   
As we shall see,  this has important consequences for the expectation values of the moments.

\subsection{Intermediate Coupling}
\label{sec:IC}

In order to assess the importance of intermediate coupling, we show in Table~\ref{tab:1} the calculated atomic Hartree-Fock (HF) values of the radial parameters of the Slater integrals for representative elements among the various transition metal series. It is seen that the HF values of the Slater integrals $F^k$ for the different metal series are comparable in size.  In practice an empirical scaling factor is used that depends on the degree of (de)localization of the valence electrons. In localized atomic systems the Coulomb and exchange parameters typically require  a scaling to 70-80\% of the HF value to account for interactions with configurations omitted in the calculation \cite{thole85}, but fully itinerant systems might have to be scaled down to 10-20\% \cite{thole94}. In line with increasing atomic number $Z$,  the value of the spin-orbit parameter is dramatically different for each of the metal series, with the largest values found for the actinides. 
While the $LS$-coupled Hund's rule ground state is a reasonable approximation for the rare earths, this will usually not hold for the actinides \cite{cricchio08}. 

%  TABLE I

\begin{table}[t]
\caption{Comparison of the radial parameters, $F^k(\ell,\ell)$, for the Coulomb interaction and the spin-orbit interaction, $\zeta_{\ell}$, for actinides with rare earths \cite{thole85} and $3d$ transition metals \cite{vanderlaan92}. All values are in eV. The Slater integrals have been reduced to 80\% of the atomic Hartree-Fock values. }
\label{tab:1}       % Give a unique label
\begin{tabular}{p{1.5cm}p{1.5cm}p{2cm}p{2cm}p{2cm}p{2cm}}
\hline\noalign{\smallskip}
&& $F^2$ & $F^4$ & $F^6$ & $\zeta_{\ell}$ \\
\noalign{\smallskip}\svhline\noalign{\smallskip}
$^{25}$Mn$^{2+}$& $3d^5$ & 8.25 & 5.13 &  & 0.040 \\
$^{64}$Gd$^{3+}$& $4f^7$ &11.60 & 7.28 & 5.24 & 0.197 \\
$^{96}$Cm$^{3+}$ &$5f^7$ & 8.37 & 5.46 &  4.01 & 0.386 \\
\noalign{\smallskip}\hline\noalign{\smallskip}
\end{tabular}
\end{table}

To express the wavefunctions in intermediate coupling, where $\{\zeta_{\ell}, F^k, G^k \} \neq 0$,  we can choose either the $jj$- or $LS$-basis states, but in both cases the Hamiltonian has off-diagonal matrix elements. 
The spin-orbit interaction is diagonal in $jj$ coupling and Eq.~(\ref{eq:7}) gives $\langle {\vec{l}} \cdot {\vec{s}} \rangle$ =
$-$4, $- \frac{1}{2}$, and $3$ for $| (5/2,5/2)_4 \rangle$, $|(5/2,7/2)_4 \rangle$, and $| (7/2,7/2)_4 \rangle$, respectively, with radial part $\zeta$.
Transforming this diagonal Hamiltonian to the $LS$-coupled basis $( |^3H_4 \rangle, |^1G_4 \rangle,   |^3F_4 \rangle )$ using Eq.~(\ref{eq:13}) and then including the diagonal Coulomb interaction gives
the full Hamiltonian
\begin{equation}    % EQUATION 17
{\vec{H}}^{(LS)}_{J=4} = \left(   \begin{array}{ccc}
 E(^3H) - 3\zeta \, & -\sqrt{ \frac{10}{3} } \zeta & 0  \\
-\sqrt{ \frac{10}{3} }  \zeta & E(^1G) & \sqrt{ \frac{11}{3} \zeta} \\
0 & \sqrt{ \frac{11}{3} } & \, E(^3F) + \frac{3}{2} \zeta 
\end{array} \right) \;,
\label{eq:17}
\end{equation}

\noindent
with the energies of the Coulomb interaction  obtained from Eqs.~(\ref{eq:9}) and (\ref{eq:9a}) as
\begin{eqnarray}           % EQUATION 18
E(^3H) &=& F_0  - \frac{1}{9} F^2  - \frac{17}{363} F^4 - \frac{25}{47157} F^6  , \nonumber \\
E(^1G) &=& F_0  - \frac{2}{15} F^2  + \frac{97}{1089} F^4 + \frac{50}{4719} F^6, \nonumber  \\
E(^3F) &=& F_0  - \frac{2}{45} F^2  - \frac{1}{33} F^4 - \frac{50}{1287} F^6 \;.
\label{eq:18}
\end{eqnarray}

\noindent
As is evident from the Hamiltonian in Eq.~\ref{eq:17},  the spin-orbit operator can be written as
\begin{equation}
\vec{L} \cdot \vec{S} = L_0 S_0 + \frac{1}{2} (L_{+} S_{-} + L_{-} S_{+}) ,
\label{eq:18a}
\end{equation}

\noindent
where the step-operators $L_{\pm}$ and $S_{\mp}$ mix $LS$ states with $\Delta L =  \pm 1$ and $\Delta S = \mp 1$.  

%  TABLE 2
    
\begin{table}[t]
\caption{Expectation values for the spin-orbit interaction and the orbital and spin magnetic moments for the $LS$-, $jj$-, and intermediate-coupled (IC) states of the  atomic configurations $f^2$ and $f^5$.   For IC, the Slater integrals for the actinides were reduced to 80\% of the Hartree-Fock values.
Also given are the occupation numbers $n_{5/2}$ and $n_{7/2}$, which are related to $n$ and $\langle {\vec{l}} \cdot {\vec{s}} \rangle$ by Eqs.~(\ref{eq:25}) and (\ref{eq:26}).}
\label{tab:2}  
\begin{tabular}{p{0.5cm}p{0.8cm}p{1.6cm}p{1.6cm}p{1.6cm}p{1.6cm}p{1.6cm}p{1.4cm}}
\hline\noalign{\smallskip}
& & & $\langle {\vec{l}} \cdot {\vec{s}} \rangle$ & $n_{5/2}$ &  $n_{7/2}$ & $\langle L_z \rangle$ & $\langle S_z \rangle$ \\ 
\noalign{\smallskip}\svhline\noalign{\smallskip}
$f^2$ & $LS$ & $|^3H_4  \rangle$ & $-$3 & 1.71 & 0.29 & $-$4.8 &  0.8 \\
  &   & $|^1G_4 \rangle$ & 0 & 0.86 & 1.14 & $-$4 &  0 \\
    &   & $|^3F_4 \rangle$ & 1.5 & 0.43 & 1.57 & $-$3 & $-$1 \\
& $jj$ &   $|(\frac{5}{2},\frac{5}{2})_4 \rangle$ &  $-$4 &  2 & 0 &  $-$4.57 &  0.57  \\
&  &   $|(\frac{5}{2},\frac{7}{2})_4 \rangle$  & $-$0.5 & 1 & 1 & $-$3.8 &  $-$0.2  \\
&  &   $|(\frac{7}{2},\frac{7}{2})_4 \rangle$  & 3 & 0 & 2 & $-$3.43 &  $-$0.57  \\  
& IC & & $-$3.88 & 1.97 & 0.03 & $-$4.70 &  0.70   \\ \\
$f^5$ & $LS$ & $|^6H_{5/2} \rangle$ & $-$3 & 3 & 2 & $-$4.29 &  1.79 \\
& IC & &  $-$7.66 & 4.33 & 0.67 & $-$3.89 &  1.38  \\
& $jj$ &   & $-$10 & 5 & 0 & $-$2.86 &  0.36  \\
\noalign{\smallskip}\hline\noalign{\smallskip}
\end{tabular}
\end{table}

\subsection{Moments for $f^2$}
\label{sec:moments}

From the wavefunction coefficients we can work out the expectation value of the different moments. 

\subsubsection{Spin-Orbit Expectation Value}

For the $LS$-coupled states  $|^3H_4  \rangle$, $|^1G_4 \rangle$,  $|^3F_4 \rangle$ we obtain from Eq.~(\ref{eq:17})   that $\langle {\vec{l}} \cdot {\vec{s}} \rangle$ = $-$3, 0,  $\frac{3}{2}$, respectively.
These results can be directly verified using Eq.~(\ref{eq:5}), which is valid  for pure $LSJ$ states.

Substituting the values of the Slater integrals $F^k$ and spin-orbit interaction $\zeta$,  and diagonalizing the Hamiltonian, we obtain the wavefunction coefficients in intermediate coupling, and from those we obtain the expectation value $\langle {\vec{l}} \cdot {\vec{s}} \rangle$. The result is shown in Table~\ref{tab:2} and can be compared with the values for $LS$ and $jj$-coupled states.

Alternatively, writing the wavefunction  in $jj$-coupled states
we can use the general expression in Eq.~(\ref{eq:7}). 
An arbitrary state of the  $\ell^2$ configuration can be written as
\begin{equation}         % EQUATION 19
\psi (\ell^2)=c_{11} |j_1,j_1 \rangle + c_{12} | j_1,j_2 \rangle + c_{22} |j_2,j_2 \rangle,
\label{eq:19}
\end{equation}

\noindent
where the wave-function coefficients fulfill the conditions $n_{j_1}=2c^2_{11}+c^2_{12}$ and $n_{j_2}=c^2_{12}+2c^2_{22}$.
For instance, using the expression of $|^3H_4 \rangle$ in Eq.~(\ref{eq:16}) gives $n_{5/2}$ = 1.714 and $n_{7/2}$ = 0.286, so that with Eq.~(\ref{eq:7}) we obtain $\langle {\mathbf{ l}} \cdot {\vec{s}} \rangle = -2 n_{j_1} + \frac{3}{2} n_{j_2} = -3$.

\subsubsection{Orbital and Spin Magnetic Moments}

To obtain the magnetic moments, we assume here that the magnetic field is infinitely small, so that $J$ is a good quantum number. Furthermore, we take the magnetic ground state at $T$ = 0 as $M_J=-J$.

We write $|LSJM_J \rangle$ as a sum over $|LSM_L,M_S \rangle$  using the Clebsch-Gordan coefficients 
\begin{equation}     % EQUATION 22
\langle   LSM_LM_S  |  J M_J \rangle = (-1)^{L-S-M_J}   [J]^{\frac{1}{2}}
\left( \begin{array}{ccc}
J & L & S  \\
M_J & M_L & M_S  
\end{array} \right)\; .
\label{eq:22}
\end{equation}

\noindent
The non-zero values for $f^2(J=4)$ are listed in Table~\ref{tab:3}. Only if the moments are stretched, i.e., $L+S=J$,
the $|LSJM_J \rangle$  state is a single determinant.

The expectation values of the orbital and spin magnetic moment are
\begin{eqnarray}     % EQUATION 20  and 21
\langle L_z \rangle &\equiv & \langle LSJM_J | L_z | LSJM_J \rangle 
= \sum_{M_L} \left| \langle  LSM_LM_S |  JM_J \rangle \right| ^2 M_L \;, 
\label{eq:20}   \\
\langle S_z \rangle & \equiv & \langle  LSJM_J  | S_z |  LSJ M_J \rangle  
= \sum_{M_S}  \left| \langle  LSM_LM_S  |  J M_J \rangle \right| ^2 M_S \;. 
\label{eq:21}
\end{eqnarray}
%

%  TABLE 3

\begin{table}[t]
\caption{Values of  non-zero Clebsch-Gordan coefficients $\langle  LSM_LM_S  |  J M_J \rangle$ with  $ \left|  J M_J \right\rangle$ =  $\left|  4,-4 \right\rangle$.}
\label{tab:3}       % Give a unique label
\begin{tabular}{p{2cm}p{4cm}p{4cm}}
\hline\noalign{\smallskip}
$^{2S+1}L_J$          & $ \langle  LSM_LM_S  |  J M_J \rangle$ &  value\\
\noalign{\smallskip}\svhline\noalign{\smallskip}
$^3H_4$ & $ \langle  5,1,-5,1  |  4,-4 \rangle$  & $\frac{3}{\sqrt{11}}$  \\
                  & $ \langle  5,1,-4,0  |  4,-4 \rangle$  & $-\frac{3}{\sqrt{55}}$  \\
                  & $ \langle  5,1,-3,-1  |  4,-4 \rangle$  & $\frac{1}{\sqrt{55}}$  \\
$^1G_4$ & $ \langle  4,0,-4,0 |  4,-4 \rangle$  & 1 \\
$^3F_4$ & $ \langle  3,1,-3,-1  |  4,-4 \rangle$  & 1  \\
\noalign{\smallskip}\hline\noalign{\smallskip}
\end{tabular}
\end{table}

\noindent
The obtained diagonal elements of $\langle L_z \rangle$ and $\langle S_z \rangle$ for the $LS$-coupled states are given in Table~\ref{tab:2}.
Switching to $jj$-coupled basis states using Eq.~(\ref{eq:13}) gives
\begin{equation}   % EQUATION 23
\langle L_z \rangle ^{(jj)} =
\left( \begin{array}{ccc} 
-4.57 & -0.42 &  0 \\
-0.42 &  -3.8 & 0.54 \\
0 &   0.54 &  -3.43
\end{array}  \right) \;,  \,\,\,\,
\langle S_z \rangle ^{(jj)} =
\left( \begin{array}{ccc} 
0.57 &  0.42 &  0 \\
 0.42 &  -0.2 & -0.54 \\
0 &   -0.54 &  -0.57
\end{array}  \right) \;. 
\label{eq:23}
\end{equation}

\noindent
Obviously, the trace of  both $\langle L_z \rangle$ and $\langle S_z \rangle$ is conserved in different coupling schemes.  Furthermore,  the matrix $ \langle L_z \rangle + \langle S_z \rangle = \langle J_z \rangle $ has diagonal elements equal to $-$4, while non-diagonal elements vanish.

Results for the other $f^n$ configurations can be found in Ref.~\cite{vanderlaan96}, which
shows that the type of coupling has an important influence on the values of the moments.
In Table~\ref{tab:2}, we reproduce the values of  the moments for $f^5  (J = 5/2)$. 
If Pu would be ferro- or ferrimagnetic, then $\langle L_z \rangle$ and $\langle S_z \rangle$ could be obtained from x-ray magnetic circular dichroism (XMCD). An alternative is to measure $\langle {\vec{l}} \cdot  {\vec{s}} \rangle$, which gives conclusive information about the type of angular coupling in the material. It is clear from from Table~\ref{tab:2} that there is a huge variation in  $\langle {\vec{l}} \cdot  {\vec{s}} \rangle$ across the three coupling cases. We will show in Sec.~\ref{sec:sumrule} how to extract the expectation value of the spin-orbit interaction from the EELS or XAS branching ratio.

\section{Spectral Calculations}
\label{sec:spectral}

We briefly mention how the multiplet calculations are performed.
Multiplet theory \cite{vanderlaan06} is ideally suited to calculate the core-level spectra for EELS or XAS at the $N_{4,5}$ and $O_{4,5}$ edges. These calculations for the transitions $5f^n \to d^9 5f^{n+1}$are performed in the same way as for the $M_{4,5}$ and $N_{4,5}$ absorption edges of the rare earths \cite{thole85}, using only different radial parameters of the spin-orbit and Slater integrals (see Table~\ref{tab:1}). Contrary to band-structure calculations, the multiplet structure is calculated in intermediate coupling, which treats spin-orbit, Coulomb, and exchange interactions on equal footing. 

First, the initial and final state wavefunctions are calculated in intermediate coupling using the atomic Hartree-Fock method with relativistic correction  \cite{cowan68,cowan81}. The electric-multipole transition matrix elements are calculated from the initial state to the final state levels of the specified configurations. At low photon energies below a few keV only electric-dipole transitions play a role. The electric-dipole selection rules from the ground state strongly limit the number of accessible final states, so that compared to the total manifold of final states, the allowed transitions are located within narrow energy regions \cite{thole85}, historically called white lines due to their appearance on photographic plates.

Examples for the $5f^0 \to d^95f^1$ transitions of the actinide $O_{4,5}$, $N_{4,5}$, and $M_{4,5}$ edges, corresponding to the $5d$, $4d$ and $3d$ core levels, respectively, as well as numerical results for other $f^n$ configurations can be found in Ref.~\cite{moore21}.

\section{Spin-Orbit Interaction and Sum Rule Analysis}
\label{sec:sumrule}

The spin-orbit sum rule relates the angular dependent part of the spin-orbit interaction to the EELS or XAS branching ratio, i.e., the intensity ratio of the core $d$ spin-orbit split $j$-manifolds in the $f^n \to d^9f^{n+1}$ transition. This is discussed in detail in this section and the validity of the sum rules is discussed in the next section.

While the expectation value of the orbital and spin magnetic moments, $\langle L_z \rangle$ and $\langle S_z \rangle$, can only be measured with XMCD, the spin-orbit interaction, $ \langle {\vec{l}} \cdot {\vec{s}} \rangle$,   can be obtained from the branching ratio of the XAS or EELS spectra. 

For the $f$ shell, the number of electrons, $n_f$, and the expectation value of the angular part of the spin-orbit interaction, $\langle w^{110} \rangle$, is given by
\begin{eqnarray}        % EQUATION 25  and 26
n_f &=& n_{7/2} + n_{5/2} \;,
\label{eq:25}     \\
\langle w^{110} \rangle &\equiv& \frac{2}{3} \left\langle {\vec{l}} \cdot {\vec{s}} \right\rangle = n_{7/2} - \frac{4}{3} n_{5/2} \;,
\label{eq:26}
\end{eqnarray}

\noindent
where $n_{7/2}$ and $n_{5/2}$ are the electron occupation numbers of the angular-momentum levels $j$ = 7/2 and $j$ = 5/2 of the $f$ shell. 

A {\it{sum rule}} \cite{thole88b,thole88l,vanderlaan88,vanderlaan96}
relates the expectation value of the angular part of the $5f$ spin-orbit interaction per hole to the branching ratio, $B$, of the core $d$ to valence $f$ transitions in EELS or XAS,
\begin{equation}             % EQUATION 27
\frac{\langle w^{110} \rangle}{14-n_f} -\Delta = - \frac{5}{2} \left( B - \frac{3}{5} \right) \;,
\label{eq:27}
\end{equation}
	     
\noindent
where the branching ratio for the $N_{4,5}$ edge is defined as 
\begin{equation}               % EQUATION 28
B \equiv \frac{I(N_5)}{I(N_5)+I(N_4)},
\label{eq:28}
\end{equation}

\noindent
with $I(N_5)$ and $I(N_4)$ the integrated intensities of the $N_5$ $(4d_{5/2} \to 5f_{5/2,7/2})$ and $N_4$ $(4d_{3/2} \to 5f_{5/2})$ peaks. Since the expectation value of $w^{110}$ is for the angular part of  the spin-orbit interaction, it does not include the radial part, which is approximately constant for given element. Therefore, the spin-orbit sum-rule analysis is complementary to e.g., optical spectroscopy, where energy level separations are measured. Thus the sum rule analysis reveals the proper angular-momentum coupling scheme in the material.

\begin{figure}[t]           %FIGURE       N-calc            FIG. 6 
\sidecaption[t]
\includegraphics[scale=0.2]{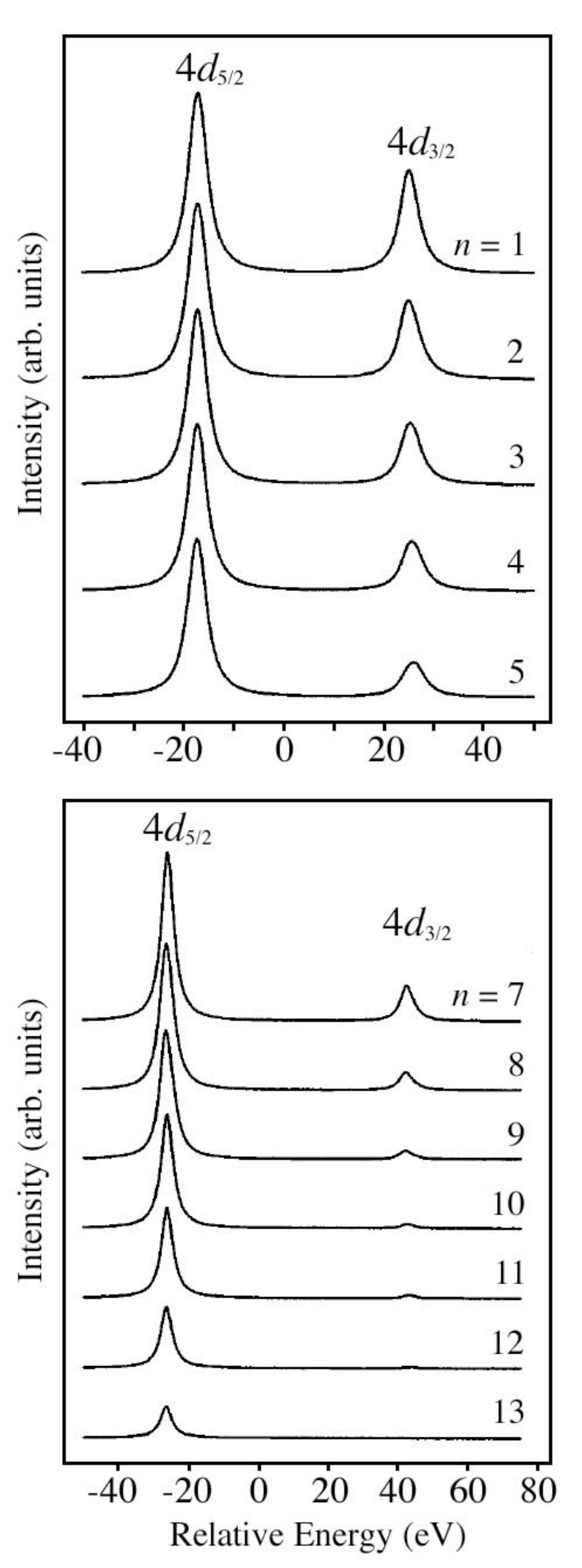}
\caption{ 
Calculated electric-dipole transitions for the $N_{4,5}$ EELS or XAS spectra using many-electron atomic theory in intermediate coupling for $^{92}$U $f^ 1$ to $f^5$ and $^{100}$Fm $f^7$ to $f^{13}$. The convolution by 2 eV corresponds to the intrinsic lifetime broadening. The spectrum for $f^6$ (not shown) has only a $4d_{5/2}$ peak but no $4d_{3/2}$ peak. From Ref.~\cite{vanderlaan04}. 
The calculated spectra show very good agreement with the experimental $N_{4,5}$ spectra in Fig.~\ref{N-exp}, where
$n$(Th) $\approx$ 1, $n$(U) $\approx$ 3, $n$(Np) $\approx$ 4, $n$(Pu) $\approx$ 5, $n$(Am) $\approx$ 6, and $n$(Cm) $\approx$ 7.
}
\label{N-calc}
\end{figure}            % End Figure

\begin{figure}[t]           %FIGURE    N-exp           FIG. 5
\sidecaption[t]
\includegraphics[scale=0.3]{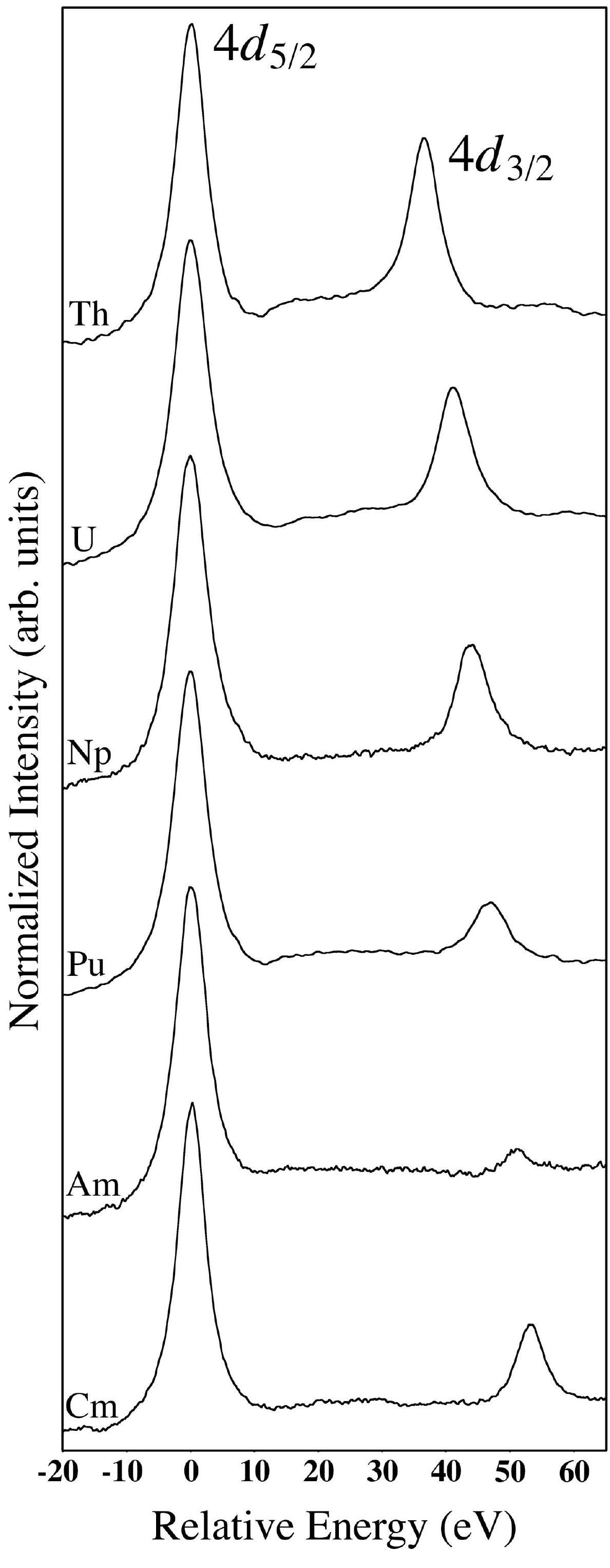}
\caption{ 
Experimental $N_{4,5}$  EELS spectra for the $\alpha$-phases of Th, U, Np, Pu, Am, and Cm metal, normalized to the $N_5$ peak height. It is observed that the intensity of the $N_4$ ($4d_{5/2}$) peak gradually decreases in intensity relative to the $N_5$ ($4d_{3/2}$) peak going from Th to Am, then increases again for Cm. From Refs.~\cite{moore07a,moore07b}. The branching ratio of these spectra gives direct information about the expectation value of the $5f$ spin-orbit  interaction in the actinde metal ground state.}
\label{N-exp}
\end{figure}

Figure~\ref{N-calc} shows the calculated electric-dipole transitions $4d^{10}5f^n \to 4d^95f^{n+1}$ for the $N_{4,5}$ EELS or XAS spectra using many-electron atomic theory in intermediate coupling for $^{92}$U $5f^1$ to $f^5$ and $^{100}$Fm $5f^7$ to $f^{13}$ (Ref.~\cite{vanderlaan96}). The line spectra are convoluted by 2 eV, which corresponds to the intrinsic lifetime broadening. The spectrum for $f^6$ (not shown) contains only a $N_5$ peak while the $N_4$ peak is zero. The calculated results in Fig.~\ref{N-calc} are in very good agreement with the experimental results shown in Fig.~\ref{N-exp}.

            % End Figure

\section{Validity of the Sum Rule}
\label{sec:validity}

\begin{figure}[t]           %FIGURE         SumRule           FIG. 7
\sidecaption[t]
\includegraphics[scale=0.32]{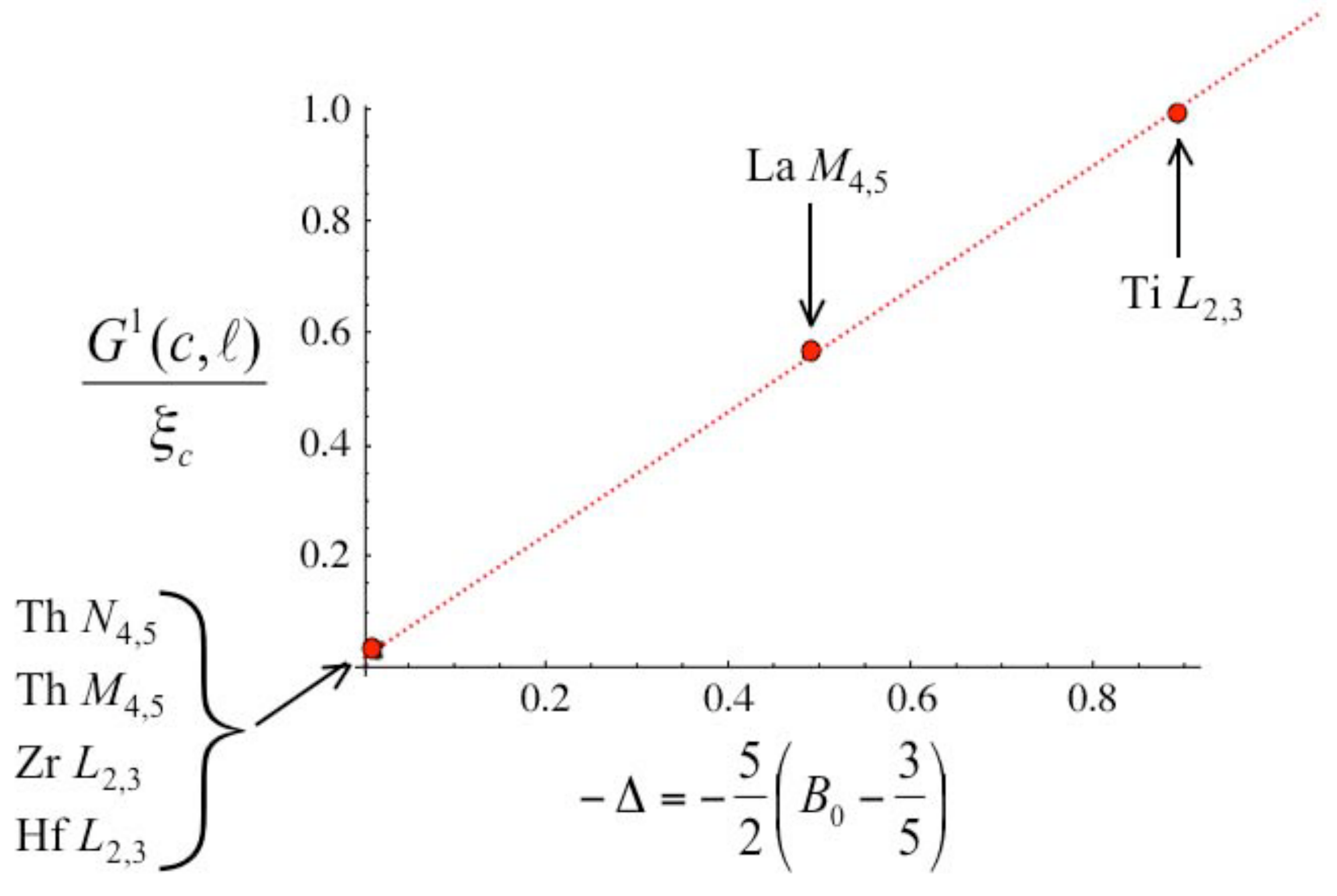}
\caption{ 
Validity of the spin-orbit sum rule. Theoretical values obtained using relativistic atomic Hartree-Fock calculations show that the correction factor, $\Delta$, in Eq.~(\ref{eq:27}) is proportional to the ratio of the core-valence exchange interaction and the core spin-orbit interaction, $G^1(c,\ell)/ \zeta_c$. The condition for the sum rule is that the total angular momentum of the core level, $j = c \pm \frac{1}{2}$ is a good quantum number, i.e., there is no mixing of the $j$ = 5/2 and $j$ = 3/2 core levels. This means the core-valence exchange interaction should be much smaller than the core spin-orbit interaction [$G^1(c,\ell)\ll \zeta_c$]. Numerical values are given in Ref.~\cite{vanderlaan04}.        }
\label{SumRule}
\end{figure}              % End Figure

The sum rule in Eq.~(\ref{eq:27}) contains a correction term $\Delta$, which can be calculated using Cowan's relativistic Hartree-Fock code \cite{cowan68,cowan81}. In the same way as the spin magnetic moment sum rule in x-ray magnetic circular dichroism (XMCD) \cite{vanderlaan98} the sum rule in Eq.~(\ref{eq:27}) is strictly only valid in the absence of core-valence electrostatic interactions, or so called $jj$ mixing, in which case the correction term $\Delta$  becomes zero \cite{vanderlaan04}. In the Periodic Table, the different transition-metal series show a different behavior due to diverse valence-electron interaction. For $3d$ transition metals, the application of the spin-orbit sum rule for the $L_{2,3}$ branching ratio is severely hampered by the large $2p$-$3d$ exchange interaction that is of similar size as the $2p$ spin-orbit interaction \cite{thole88b}.  The same is true for the $M_{4,5}$ edges of the lanthanides, where the $3d$-$4f$ exchange interaction is quite strong compared to the $3d$ core spin-orbit interaction \cite{thole85}.  On the other hand, for the $L_{2,3}$ edges of $4d$ and $5d$ transition metals, associated with the deep $2p$ core level, the sum rule is expected to hold quite well \cite{thole88b}.  
Theoretical values obtained using relativistic atomic Hartree-Fock calculation (Cowan code) in Fig.~\ref{SumRule} show that the correction factor, $\Delta$, is proportional to the ratio of the core-valence exchange interaction and the core spin-orbit interaction, i.e., $G^1(c,\ell)/ \zeta_c$. Thus the condition for the sum rule is that the total angular momentum of the core hole, $j = c \pm \frac{1}{2}$ is a good quantum number, in other words there should be no mixing of the $j$ = 5/2 and $j$ = 3/2 core states. This corresponds to the core-valence exchange interaction much smaller than the core spin-orbit interaction, $G^1(c,\ell) \ll \zeta_c$ \cite{vanderlaan04}.

Even in the worst case of the rare earth $M_{4,5}$ edges, the trend in the branching ratio can still be used to obtain the relative population of spin-orbit split states, as was demonstrated for Ce systems \cite{vanderlaan86}. The situation is favorable for the $M_{4,5}$ and $N_{4,5}$ edges of the actinides, given the small exchange interactions between the $3d$ and $4d$ core levels and the $5f$ valence states.  This means that the EELS and XAS branching ratios depend almost solely on the $5f$ spin-orbit expectation value per hole, thus affording an unambiguous probe for the $5f$ spin-orbit interaction in actinide materials.

\section{Experimental Results for the $N_{4,5}$ Edges}
\label{sec:N45}

The experimental $N_{4,5}$ $(4d \to 5f )$ EELS spectra for $\alpha$-phase Th, U, Np, Pu, Am, and Cm metals are displayed in Fig.~\ref{N-exp}. All spectra are normalized to the $N_5$ peak height. Noticeable is the gradually growing separation between the $N_4$ and $N_5$ peaks from Th to Cm, in pace with the increase in $4d$ spin-orbit splitting with atomic number.  Second and more importantly, the intensity of the $N_4$ $(4d_{5/2})$ peak gradually decreases in intensity relative to the $N_5$ $(4d_{3/2})$ peak going from Th to Am, then abruptly increases for Cm \cite{moore07a,moore07b}. Applying the sum-rule analysis to the experimental branching ratio gives the values of the $5f$ spin-orbit interaction per hole.

\begin{figure}[t]           %FIGURE        SO-result           FIG. 8
\sidecaption[t]
\includegraphics[scale=0.7]{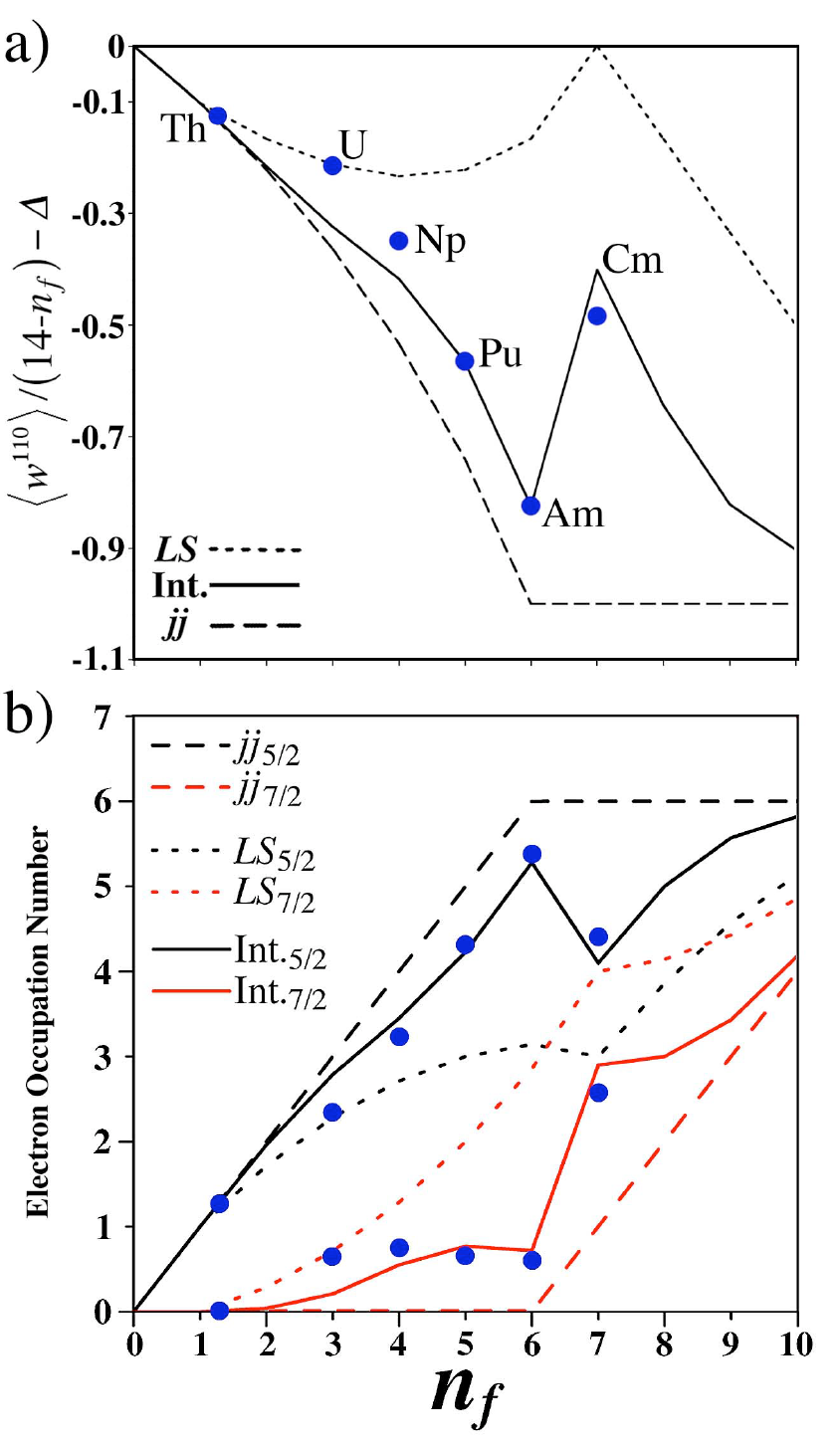}
\caption{ 
(a) Ground-state spin-orbit interaction per hole as a function of the number of $5f$ electrons ($n_f$). The three theoretical angular-momentum coupling schemes are shown ($LS$, $jj$, and intermediate coupling). The dots give the results of the spin-orbit sum-rule analysis using the experimental $N_{4,5}$ branching ratio of each metal in Fig.~\ref{N-exp}. (b) The spin-orbit interaction values are converted to the electron occupation numbers $n_{5/2}$ and $n_{7/2}$ of the $j$ = 5/2 and 7/2 levels in the $5f$ state. Shown are the three theoretical coupling schemes (curves for $LS$, $jj$, and intermediate coupling) together with the dots representing the experimental EELS results for each metal. \cite{moore07a,moore07b}    }
\label{SO-result}
\end{figure}

In order to visualize the spin-orbit analysis of the EELS spectra in relation to the results of our atomic calculations, both are shown as a plot of $\langle w^{110} \rangle / (14-n_f) - \Delta$ versus the number of $5f$ electrons in Fig.~\ref{SO-result}(a). The curves for the three theoretical angular-momentum coupling schemes, $LS$, $jj$, and intermediate, as calculated using a many-electron atomic model, are plotted as a short-dashed, long-dashed, and solid line, respectively.  The EELS results are indicated by the blue points.  Thorium metal falls on all three curves, due to the fact that it takes two electrons to tangle and with less than one $5f$ electron in Th there is no difference between the coupling mechanisms. U falls directly on the $LS$-coupling curve, Np between the $LS$ and intermediate curve, and Pu, Am, and Cm all fall on or near the intermediate coupling curve.  The intermediate curve is strongly shifted towards the $jj$ limit for Pu and Am, evidence of the strong preference of the $5f$ electrons to occupy the $j=5/2$ level in both metals.  However, for Cm there is a sudden and pronounced shift in the intermediate coupling curve toward the $LS$ limit.

            % End Figure

The values of $n_f$ and $\langle w^{110} \rangle$ can be converted into the electron occupation numbers $n_{7/2}$ and $n_{5/2}$ using Eqs.~(\ref{eq:25}) and (\ref{eq:26}). Note that this is just a different way of presenting the same data. The experimental and theoretical results are displayed in Fig.~\ref{SO-result}(b), where the number of electrons in the $j=5/2$ and $j=7/2$ levels as calculated in intermediate coupling using the atomic model are drawn with black and red lines, respectively. Again, the experimental EELS results are indicated with blue points.  Apart from the slight deviation in the lighter actinides, U and Np, which is caused by delocalization of the $5f$ states and thus indicates a departure from the atomic model, the EELS results are in excellent agreement with the theoretical curves.  Figure~\ref{SO-result}(b) clearly shows that for the actinide metals up to and including Am, the $5f$ electrons strongly prefer the $f_{5/2}$ level.  However, this changes in a striking manner at Cm, where not only does the electron occupation sharply increases for the $f_{7/2}$ level, but even {\it{decreases}} for the $f_{5/2}$ level.

The physical origin of the abrupt and striking change in the values between Am to Cm can be understood from the angular momentum coupling. In $jj$ coupling the electrons prefer to be in the $f_{5/2}$ level, which however can hold no more than six, so that the maximal energy gain in $jj$ coupling is obtained for Am $f^6$, where the $f_{5/2}$ level is full.  However, for Cm $f^7$ at least one electron will be relegated to the $f_{7/2}$ level.  The $f^7$ configuration has the maximal energy stabilization due to the exchange interaction, with parallel spin in the half-filled shell, which can only be achieved in $LS$ coupling. Thus the large changes observed in the electronic and magnetic properties of the actinides at Cm are due to the transition from optimal spin-orbit stabilization for $f ^6$ to optimal exchange stabilization for $f^7$.  In all cases, spin-orbit and exchange interaction compete with each other, resulting in intermediate coupling; however, increasing the $f$-count from 6 to 7 shows a clear and pronounced shift in the power balance in favor of the exchange interaction. The effect is in fact so strong that, compared to Am, not one but two electrons are transferred to the $f_{7/2}$ level in Cm [Fig.~\ref{SO-result}(b)]. Therefore, in Cm metal, the angular-momentum coupling in the $5f$ states plays a decisive role in the formation of the magnetic moment, with Hund's rule coupling being the key to producing the large spin polarization that dictates the newly found crystal structure of Cm under pressure \cite{heathman05,moore07a,soderlind08a}. 

\subsection{What Our Results Mean for Pu Theory}

The spin-orbit sum rule suggests an $f$-count near 5 for Pu, with 5.4 being a reasonable upper limit \cite{moore06a,moore07a,moore07b}. Further evidence for this $f$-count comes from Anderson impurity calculations for Pu \cite{vanderlaanunp} that explain photoemission results on 1 to 9 monolayers thin films of Pu metal \cite{gouder01,havela02}. The $4f$ core-level photoemission spectra display a screened and unscreened peak, thereby acting as a ruler for the degree of localization. The results for the $f$-count are in agreement with recent DMFT calculations by Shim {\it{et al.}} \cite{shim08} and Marianetti {\it{et al.}} \cite{marianetti08}, as well as LDA+U calculations by Shick {\it{et al.}} \cite{shick07}, which explain the three-peak structure in $5f$ photoemission and the relatively high electronic specific heat. The absence of experimentally observed magnetic moments in any of the six allotropic phases of Pu metal \cite{lashley05} is thought to be due to Kondo screening and a non-single Slater-determinant ground state in these respective models. The lack of local moments $\delta$-Pu has also been recently explained theoretically in terms of electron coherence using dynamic mean-field theory \cite{marianetti08}. 
Using DFT, a model where spin-orbit interactions and orbital polarization are strong, but spin polarization is zero has been employed to yield a nonmagnetic configuration of $\delta$-Pu~\cite{soderlind08b}. This has received ancillary support  by polarized neutron scattering measurements on single crystal PuCoGa$_5$, which shows the orbital moment dominates the magnetization~\cite{hiess08}.
Recent magnetic susceptibility measurements have shown that local magnetic moments in the order of 0.05 $\mu_{\mathrm{B}}$/atom form in Pu as damage accumulates due to self-irradiation \cite{mccall06}.

\section{Conclusions}
\label{conclusions}

The rare-earth metals have localized and atomic-like $4f$ states across the series, leading to strong magnetic moments. The $5d$ transition metals are itinerant and band-like, behaving as typical metals with wide bands that strongly participate in bonding. The $5f$ actinide metals exhibit a behavior similar as the $5d$ transition metals in the early actinides Th, Pa, U, and Np, but then as the $4f$ rare earths for the middle actinides Am, Cm, Bk, and Cf. This is directly attributed to a transition from itinerant to localized $5f$ states that occurs at Pu. This transition can be examined through the $N_{4,5}$ EELS spectra and the spin-orbit sum-rule analysis. The relative $N_4$ peak intensity reduces up to Am, then increases for Cm in EELS spectra, where the $N_4$ and $N_5$ peaks are only marginally broadened by multiplet structure.  Across the actinide series, we see the light metals exhibit $LS$ coupling while the middle metals Pu, Am, and Cm exhibit intermediate coupling.  It is the transition from $LS$ to intermediate coupling which reveals the transition from itinerant to localized $5f$ states. This means that the EELS  $N_{4,5}$ spectra and spin-orbit sum-rule analysis can be used as a measure of the degree of itinerancy in the actinide $5f$ states.

The EELS spectra and spin-orbit analysis clearly support that $n_f  \approx 5$ and not 6 in Pu and that it falls near intermediate coupling curve. The 5$f^5$ configuration with a spin-orbit interaction that adheres to intermediate coupling near the $jj$ limit  begs the question of why Pu is not magnetic. Recent advances in band theory have begun to address this question, such as DFT with zero spin polarization but strong spin-orbit and orbital polarization \cite{soderlind08b}, DMFT calculations by Shim {\it{et al.}} \cite{shim08} and Marianetti {\it{et al.}} \cite{marianetti08}, and LDA+U calculations by Shick {\it{et al.}} \cite{shick07}

Americium falls precisely on intermediate coupling curve and has the largest spin-orbit interaction of all actinide metals. The results anchor the $f$-count for the adjacent actinides and clearly show the metal is non-magnetic because the $5f^6$ has a ground state $J=0$. Finally, the results for Cm prove the spin-orbit sum rule works, showing the intermediate coupling curve bends back to $LS$ curve. The experimental results \cite{moore07a} show that the angular-momentum coupling mechanism dictates large spin polarization and explain the magnetic stabilization of Cm observed by Heathman {\it {et al.}} \cite{heathman05}

\begin{acknowledgement}
We like to thank Wolfgang Felsch for his valuable contribution on the Ce comparison. Part of this work was performed under the auspices of the U.S. Department of Energy by Lawrence Livermore National Laboratory.
\end{acknowledgement}

% BibTeX users please use
% \bibliographystyle{}
% \bibliography{}

%\bibliographystyle{apsrev}
%%%%%%\bibliographystyle{plain}
%%%%%%%%\bibliography{Ac} % Produces the bibliography via BibTeX.

\end{document}